# A lipidated peptide derived from the C-terminal tail of the vasopressin 2 receptor shows promise as a new β-arrestin inhibitor

Rebecca L. Brouillette[a,b], Christine E. Mona[c,d], Michael Desgagné[a,b], Malihe Hassanzedeh[a,b], Émile Breault[a,b], Frédérique Lussier[a,b], Karine Belleville[a,b], Jean-Michel Longpré[a,b], Michel Grandbois[a,b], Pierre-Luc Boudreault[a,b,e], Élie Besserer-Offroy[a,e,f,g,#,*], Philippe Sarret[a,b,e,#*]

[a]Department of Pharmacology-Physiology, Faculty of Medicine and Health Sciences, Université de Sherbrooke, Sherbrooke, QC, Canada.

[b]Institut de Pharmacologie de Sherbrooke, Université de Sherbrooke, Sherbrooke, QC, Canada.

[c]Ahmanson Translational Theranostics Division, Department of Molecular and Medical Pharmacology, David Geffen School of Medicine, University of California-Los Angeles, Los Angeles, CA, USA.

[d]Jonsson Comprehensive Cancer Center, UCLA Health, Los Angeles, CA, USA.

[e]RECITAL International Partnership Lab, Université de Caen-Normandie, Caen, France & Université de Sherbrooke, Sherbrooke, QC, Canada.

[f]Université de Caen Normandie, INSERM U1086 – Anticipe, Normandie Université, Caen, France.

[g]Baclesse Comprehensive Cancer Center, UNICANCER, Caen, France.

[#]Lead Authors

***Co-Corresponding Authors**

**Philippe Sarret, Ph.D.**
Department of Pharmacology-Physiology
Faculty of Medicine and Health Sciences
Université de Sherbrooke
3001, 12th Avenue North
Sherbrooke, QC, Canada, J1H 5N4
Ph.: +1 (819) 821-8000, Ext. 72554
Philippe.Sarret@USherbrooke.ca

**Elie Besserer-Offroy, Ph.D.**
Inserm U1086 – Anticipe
Baclesse Comprehensive Cancer Center
Université de Caen-Normandie
3 avenue du Général Harris
14076 Caen Cedex 5, France
Ph.: +33 2 31 45 50 70
Elie.Besserer-Offroy@inserm.fr

## Highlights

- ARIP is a membrane-tethered lipopeptide derived from the C-terminal tail of V2R
- ARIP inhibits β-arrestin recruitment to multiple G protein-coupled receptors
- ARIP does not recruit β-arrestins on its own nor interfere with G protein signalling
- Intrathecal administration of ARIP potentiates morphine-induced analgesia in rats
- ARIP is a promising new pharmacological tool to understand the contributions of β-arrestins

**Chemical compounds studied in this article:**

Arginine vasopressin (PubChem CID: 644077)

[Pyr1]apelin-13 (PubChem CID: 25085173)

DAMGO (PubChem CID: 5462471)

GLP-1 (1-37) (PubChem CID: 16131070)

Morphine (PubChem CID: 5288826)

**Abstract**

β-arrestins play pivotal roles in seven transmembrane receptor (7TMR) signalling and trafficking. To study their functional role in regulating specific receptor systems, current research relies mainly on genetic tools, as few pharmacological options are available. To address this issue, we designed and synthesised a novel lipidated phosphomimetic peptide inhibitor targeting β-arrestins, called ARIP, which was developed based on the C-terminal tail (A343-S371) of the vasopressin V2 receptor. As the V2R sequence has been shown to bind β-arrestins with high affinity, we added an N-terminal palmitate residue to allow membrane tethering and cell entry. Here, using $BRET^2$-based biosensors, we demonstrated the ability of ARIP to inhibit agonist-induced β-arrestin recruitment on a series of 7TMRs that includes both stable and transient β-arrestin binders, with efficiencies that dependent on receptor type. In addition, we showed that ARIP was unable to recruit β-arrestins to the cell membrane by itself, and that it did not interfere with G protein signalling. Molecular modelling studies also revealed that ARIP binds β-arrestins as does V2Rpp, the phosphorylated peptide derived from V2R, and that replacing the p-Ser and p-Thr residues of V2Rpp with Glu residues does not alter ARIP's inhibitory activity on β-arrestin recruitment. Importantly, ARIP exerted an opioid-sparing effect *in vivo*, as intrathecal injection of ARIP potentiated morphine's analgesic effect in the tail-flick test, consistent with previous findings of genetic inhibition of β-arrestins. ARIP therefore represents a promising pharmacological tool for investigating the fine-tuning roles of β-arrestins in 7TMR-driven pathophysiological processes.

## Graphical abstract

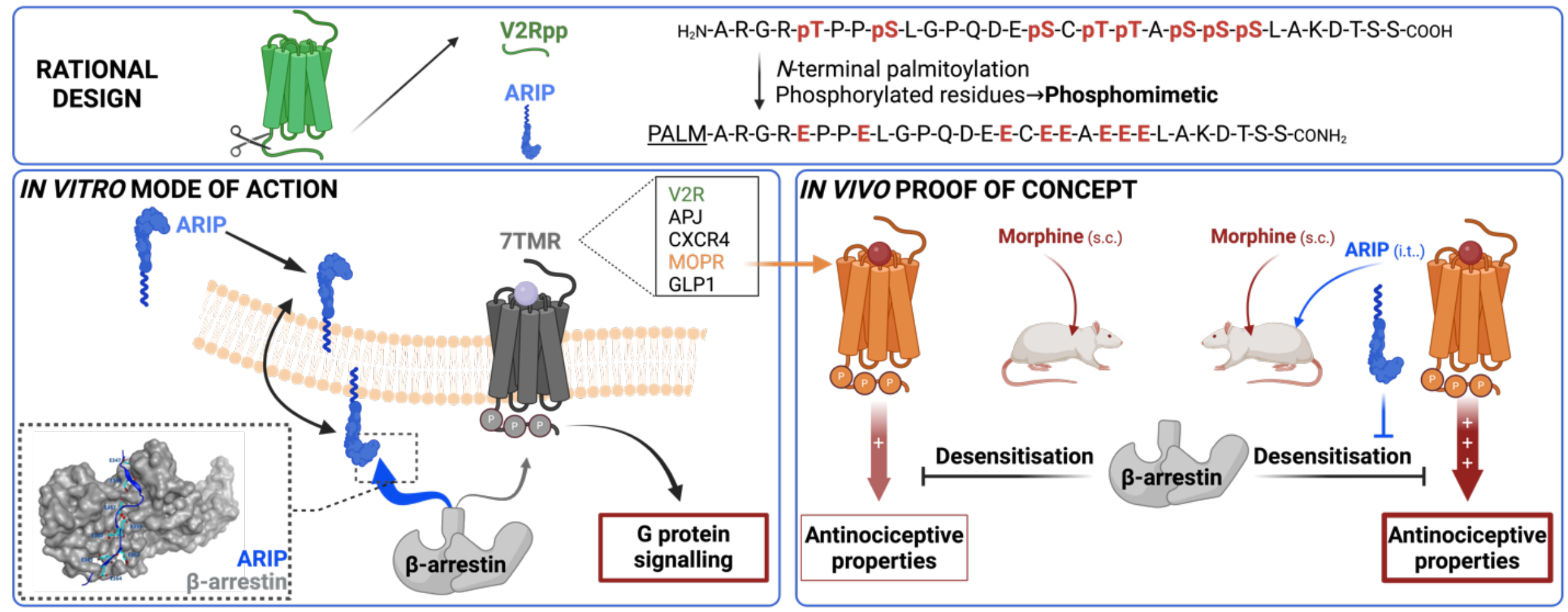

## 1. Introduction

The non-visual arrestins, identified as arrestin-2 and arrestin-3 in the order of cloning, [1] but more commonly referred to as β-arrestins 1 and 2, respectively, [2,3] are adaptor proteins that play key regulatory roles in the signalling and trafficking of seven transmembrane receptors (7TMRs).[4,5] These cell surface receptors, also known as G protein-coupled receptors (GPCRs), mediate a wide range of physiological effects and have historically been one of the most successful target classes for drug development.[6–8] Classically, β-arrestins are recruited to 7TMRs following phosphorylation of residues within the C-terminal tail and/or intracellular loops (ICL) of activated receptors. More specifically, phosphorylation of serine and threonine residues by cytosolic G protein-coupled receptor kinases (GRKs) represents a crucial step in 7TMR/β-arrestin binding. Until now, the recruitment of β-arrestins fulfils three critical cellular functions: ***(1)*** block or turn off G protein coupling, thereby causing receptor desensitisation; ***(2)*** promote receptor endocytosis by linking to clathrin, AP-2 and other endocytic adaptors to activated receptors; and ***(3)*** trigger specific downstream signalling pathways by scaffolding a wide range of proteins, including MAP kinase cascades, such as c-Raf1-MEK1-ERK1/2[9]. While β-arrestin binding is critical for regulating GPCR activity, the affinity and stability of the resulting receptor/β-arrestin complex, as well as the conformations it adopts, may differ from one couple to another. For example, certain 7TMRs have transient receptor/β-arrestin association and, after receptor internalisation, are rapidly recycled back to the cell membrane and re-sensitised (referred to here as "transient binders") while other GPCRs ("stable binders") form highly stable associations with arrestins and are retained longer within the intracellular compartments.[10–12] Additionally, 7TMR/arrestin complexes have been proposed to adopt two distinct sets of conformations: ***(1)*** the "tail" conformation, in which the arrestin primarily engages with the receptor's phosphorylated C-terminal tail, and ***(2)*** the "core" conformation, in which the β-arrestin also interacts with the receptor's intracellular core.[13] These proposed conformations, determined using chimeric receptor constructs, are believed to lead to a distinct subset of β-arrestin-associated functions. According to the "barcode" hypothesis, the conformations adopted by β-arrestins and, hence, the cellular fates enacted, are dictated by the distinct phosphorylation patterns of the 7TMRs.[14]

Understanding the actions of β-arrestins in specific receptor systems, as well as their involvement in pathophysiological states, is critical to the development of effective therapeutics targeting 7TMRs. To achieve this understanding, highly specific pharmacological tools are needed. Unfortunately, while pharmacological inhibitors directed against G proteins and/or their second messengers exist and are commercially available,[15–18] direct β-arrestin-specific inhibitors are still lacking. Blockers of the clathrin-mediated endocytosis (CME) machinery, such as the dynamin-targeting inhibitors Dynasore[19] or Dynole 34-2[20], may offer some insight; however, they affect β-arrestin-binding and non-binding proteins alike. Otherwise, the only pharmacological inhibitor reported to date that selectively targets β-arrestins is the small molecule Barbadin.[21] Barbadin interferes with the binding of β-arrestin to the β2-adaptin subunit of AP-2 and thereby blocks receptor internalisation, but does not directly inhibit arrestin recruitment or binding to 7TMRs.

Structural studies investigating the mechanism of 7TMR/β-arrestin binding have shown that V2Rpp, a phosphorylated peptide derived from the C-terminal domain of the vasopressin 2 receptor (A343 to S371), can bind β-arrestins 1 and 2.[22–24] V2Rpp has 8 phosphosites, at positions 5 (T347), 8 (S350), 15 (S357), 17 (T359), 18 (T360), 20 (S362), 21 (S363), and 22 (S364): these phospho-groups form critical contacts with the positively charged side chains of lysines and arginines in the N-terminal domain of β-arrestin.[23] Phosphosites 15, 18, 20, and 21 have been shown to be particularly important *in vitro*, as alanine mutations of the corresponding V2R residues (S357, T360, S362, S363) reduce AVP-promoted β-arrestin binding in a cellular model.[25]

In recent years, lipidated peptides derived from 7TMR intracellular domains, known as *pepducins*, have emerged as new modulators of 7TMR signalling.[26–28] Studies have shown that pepducins' lipid moiety effectively tethers the peptide to the cell membrane, where it can passively "flip-flop" between the inner and outer leaflets of the phospholipid membrane bilayer.[29,30] Once facing the cytosol, they interact with the receptor and/or its effectors and thus, modulate its signalling output. In particular, pepducins were shown to promote or inhibit agonist-mediated 7TMR signalling, even favoring bias towards specific signalling pathways.[31–35]

In the present study, we sought to develop a new β-arrestin inhibitor, called *A*rrestin *R*ecruitment *I*nhibitory *P*eptide (ARIP), by applying a similar strategy to V2Rpp. We designed a lipidated phosphomimetic peptide derived from the C-terminal tail of V2R. Using bioluminescence resonance energy transfer (BRET)-based screening assays in live cells, we demonstrated the lipopeptide's ability to inhibit agonist-induced β-arrestin recruitment against a panel of 7TMRs (V2R, CXCR4, APJ, MOPR, GLP1R), albeit with variable efficiencies. We also demonstrated that ARIP is unable to recruit β-arrestins to the cell membrane by itself and does not interfere with G protein-mediated signalling at selected receptors. Our molecular modelling studies further highlight that ARIP binds to β-arrestins with high complementarity, like V2Rpp, and that replacing the phospho-Ser and phospho-Thr residues of V2Rpp with Glu residues does not substantially affect the inhibitory activity of ARIP on β-arrestin recruitment. Finally, as an *in vivo* proof-of-concept, we found that intrathecal delivery of ARIP was effective in potentiating the analgesic effect of morphine in the acute thermal nociceptive assay (tail-flick test), a response consistent with β-arrestin inhibition.[36] Therefore, we propose ARIP as a promising new pharmacological tool for 7TMR research in drug discovery.

## 2. Materials & Methods

### *2.1. Lipopeptide synthesis, purification, and characterisation*

#### *2.1.1 ARIP*

TentaGel S RAM resin, all Fmoc-protected L-amino acids, palmitic acid, *O*-(7-Azabenzotriazol-1-yl)-*N,N,N',N'*-tetramethyluronium hexafluorophosphate (HATU), *N*,*N*-diisopropylethylamine (DIPEA), and trifluoroacetic acid (TFA) were purchased from Chem-Impex International or Matrix Innovation. All other reagents were purchased from Sigma-Aldrich or Fisher Scientific and were of the highest commercially available purity. Peptide synthesis was performed in 12 mL polypropylene cartridges with 20 µm PE frit (Applied Separations).

250 µmol of TentaGel S RAM resin (0.24 mmol/g) was deprotected with a mixture of piperidine:DMF (1:1) for 2 × 10 min. Fmoc-protected amino acids (3 eq.) were coupled using HATU (3 eq.), DIPEA (6 eq.) in DMF (10 mL) for at least 2 h. The reactant and solvent were then filtered, and the resin was washed with 10 mL DMF (2 × 5 min under agitation) and three cycles alternatively washing with 2-propanol (7 mL) or DCM (7 mL). Deprotection cycles were carried out with a mixture of piperidine:DMF (1:1) for 2 × 10 min. After deprotection, solvent was removed by filtration, the resin was washed with DMF, 2-propanol, and DCM (as described above) and the subsequent Fmoc-protected amino acid was coupled using HATU/DIPEA in DMF. Coupling of palmitate (3 eq.) was done in dry N-Methyl-2-Pyrrolidone (NMP) with HATU (3 eq.) and DIPEA (6 eq.).

The peptide was cleaved from the resin using 92.5% TFA, 2.5% TIPS, 2.5% $H_2O$, and 2.5% EDT for at least 2 hours to provide the crude mixture of the desired peptide. The resin suspension was filtered through cotton wool and the peptide precipitated in 50 mL of tert-butyl methyl ether at 0°C. The suspension was centrifuged for 20 minutes at 1,500 rpm and the pellet was resuspended in a mixture of water:acetonitrile (2:1) and lyophilized before being purified.

Purification of peptide was done on a Waters Mass-triggered preparative HPLC system (Sample Manager 2767, Binary gradient module 2545, with two 515 HPLC pump and a System Fluidics Organizer, Photodiode Array Detector 2998) paired to a SQ Detector 2 and equipped with a X Select CSH Prep $C_{18}$ (5 mm OBD 19 mm × 250 mm column) using a 25-40% gradient of acetonitrile with 0.1% formic acid in 15 min. Fractions were analyzed by UPLC/MS (Water H Class Acquity UPLC, mounted with Acquity UPLC BEH $C_{18}$ column, 1.7 µm, 2.1 mm × 50 mm and paired to a SQ Detetctor 2) using a 5–95% gradient of acetonitrile with 0.1% formic acid in 2 min. Fractions with a purity of 95% or greater were then lyophilized and stored at -20˚C until use. HRMS were performed on a Bruker MaXis 3G high resolution Q-ToF.

Purified product yielded a white powder after being freeze-dried. RP-HPLC: Retention time = 1.36 min. MS: 3439.70 (M+1), 1721.0 (M+2)/2, 1147.8 (M+3)/3, 861.1 (M+4)/4. HRMS: 1146.8902. Calculated m/z: 1146.8909. Purity at 290 nm: 97%.

*2.1.2 ARIP-4P*

The lipopeptide was synthesized using standard Fmoc solid-phase synthesis methodology with a Symphony X apparatus from Gyros Protein Technologies. The resin used is TentaGel S RAM Resin with a nominal loading at 0.25 mmol/g from Rapp Polymere GmbH. Fmoc-Ser(PO(OBzl)OH)-OH and Fmoc-Thr(PO(OBzl)OH)-OH were both purchased from Combi Blocks. All proteogenic Fmoc-protected amino acids were purchased from Combi Blocks, Chem-Impex or Matrix Innovation at the highest purity available. Other reagents (coupling reagents, base) and solvents were purchased from Chem-Impex, Matrix Innovation, or Sigma-Aldrich and were used as received. All coupling were performed using 5 eq. of amino acid (based on nominal resin loading), 5 eq. of HATU and 6 eq. of DIPEA unless otherwise stated.

After coupling a phosphorylated amino acid, two 5-min washes were performed using 3 mL of DMF containing 20 eq. of DIPEA and 18 eq. of TFA (based on the nominal loading of 0.25 mmol/g). For Fmoc deprotection of phosphorylated amino acids, a solution of 50% cyclohexylamine in DMF was used as a deprotecting solution instead of piperidine 20% to suppress the formation of β-elimination by-products.[37]

The peptide was cleaved from the resin using 2 mL of a 92.5% TFA, 2.5% TIPS, 2.5% $H_2O$, and 2.5% EDT solution for at least 3 hours to provide the crude mixture of the desired peptide. The resin suspension was filtered through cotton wool and the peptide was precipitated in 50 mL of tert-butyl methyl ether at 0°C. The suspension was centrifuged for 10 min at 3,000 rpm and the precipitate was solubilized in a mixture of water:acetonitrile (1:1) with a few drops of DMSO to ensure complete solubility. The compound was filtered and purified using a preparative HPLC-MS system, as previously described.

Purified product yielded a white powder after being freeze-dried. HRMS: 1202.1709. Calculated m/z: 1202.1705. Purity at 290 nm: 97%.

*2.2. Cell culture and transfections*

Opti-MEM was purchased from Gibco. Polyethylenimine (PEI) was purchased from Polysciences. All other tissue culture media and additives were sourced from Wisent Bioproducts.

HEK293 cells (CRL-1573 from ATCC, RRID:CVCL_0045) were cultured in high-glucose DMEM containing 10% FBS, 100 U/mL of penicillin, 100 μg/mL of streptomycin, and 2 mM of L-Glutamine. Cells were maintained at 37˚C in a humidified chamber under 5% $CO_2$. Cells were used between passages 10 and 25. BRET$^2$-based assays required transfection of cDNA plasmids, for transient expression of the recombinant proteins. The transfection procedure was as follows: 2 $\times 10^6$ cells were seeded onto 100 mm$^2$ cell culture dishes, and, 24 h later, received a total of 12 μg cDNA, prepared in Opti-MEM serum-free media along with the transfection agent polyethylenimine (PEI) at a 3:1 ratio (PEI:DNA), as previously described.[38]

*2.3. BRET$^2$ assays*

The plasmid encoding for GLP1R-RlucII was a gift from Dr. Rasmus Jorgensen, Novo Nordisk.[39] CXCR4-RlucII was a gift from Dr. Nikolaus Heveker (CR-CHU Ste-Justine)[40] and MOPR-RlucII was a gift from Dr. Louis Gendron (Université de Sherbrooke).[41] Plasmids encoding for V2R-RlucII[42], β-arrestin-1-RlucII[43] or β-arrestin-2-RlucII[44], β-arrestin-1-GFP10 or β-arrestin-2-GFP10[32], $G\alpha_{i1}$-RlucII[45], GFP10-$G\gamma_1$[46,47], GFP10-EPAC-RlucII[48] and CAAX-rGFP[42] were provided by Dr. Michel Bouvier (Université de Montréal). Plasmids for the expression of APJ, CXCR4, GLP1, MOPR, V2R, and $G\beta_1$ subunit were purchased from cDNA Resource Center (Bloomsburg University, cdna.org). Coelenterazine 400A (DeepBlue C) was sourced from GoldBio. HBSS (without $Ca^{++}$ and $Mg^{++}$) was from Gibco. White opaque 96-well plates were purchased from Falcon-Corning.

β-arrestin recruitment experiments were carried out as described previously.[49] Briefly, HEK293 cells were co-transfected with either β-arrestin-1- or -2-RlucII and 7TMR-GFP10, or with β-arrestin-1- or -2-GFP10 and 7TMR–RlucII, to measure arrestin recruitment to the receptor. Alternatively, cells were transfected with β-arrestin-1- or -2-RlucII and CAAX-rGFP, and untagged APJ as a control receptor, to measure arrestin recruitment to the plasma membrane. To monitor $G\alpha_i$ activation, HEK293 cells were co-transfected with $G\alpha_{i1}$-RlucII, GFP10-$G\gamma_2$, $G\beta_1$ and untagged APJ or CXCR4. To assess the effect of ARIP on cAMP production, cells were co transfected with GFP10-EPAC-RlucII and V2R.

24 h post-transfection, the cells were detached using trypsin (0.25%) containing 0.53 mM EDTA and seeded into white opaque 96-well plates. 48 h after transfection, cells were washed with HBSS and then incubated with fixed concentrations of ARIP for 30 min. Cells were washed with HBSS to remove free excess ARIP, then ligands or vehicle were added for 20 min in the case of the arrestin recruitment assays, and for 5 minutes in the G protein assays. Five min prior to luminescence measurement, coelenterazine-400A was added for a final concentration of 5 µM. Plate read-outs were obtained using a Berthold LB943-Mithras$^2$ plate reader where the BRET$^2$ filters were set to monitor the 515 nm/400 nm emission ratio. Time-course experiments were read at 20 s intervals with 0.5 s integration time, whereas endpoint experiments were read with a 1 s integration time.

*2.4. Animals*

All animal procedures were approved by the Ethical and Animal Care Committee of the Université de Sherbrooke (protcol number: 035-18) and were in accordance with policies and directives of the Canadian Council on Animal Care. Furthermore, all procedures involving animals followed the revised ARRIVE guidelines.[50] Adult male Sprague-Dawley rats (weighting between 250 and 300 g, Charles River Laboratories, RRID:SCR_003792) were maintained on a 12 h light/12 h dark cycle with access to food and water *ad libitum*. Rats were acclimatized to the animal facility for 4 consecutive days and to the manipulations for 3 consecutive days prior to the experimental conditions.

*2.4.1. Intrathecal administration*

Rats were lightly anesthetised with 2.5% isofluorane from Abbott Laboratories. Subsequently, a maximium volume of 25 μL of ARIP (250 nmol/kg) was administered by injection into the subarachnoid space between lumbar vertebrae L5 and L6, using a 27 G 1/2″ needle. ARIP was diluted in a vehicle composed of physiological saline, 50% DMSO and 20% polyethylene glycol 4000 (PEG4000, Sigma-Aldrich). Control animals were injected with the vehicle alone. At these doses, no sedation or visible side effects were observed.

*2.4.2. Acute pain model (tail-flick test)*

ARIP's effect on morphine-mediated antinociception at low doses was assessed using an acute thermal nociceptive test (tail-flick test). This test measures the latency (in seconds) for a rat to withdraw its tail from an acute nociceptive stimulus, a water bath maintained at 52 °C. The effects of ARIP and the vehicle were assessed 12 h following i.t. injection. The thermal threshold latencies were determined at baseline (before morphine injection) and at 5, 20, 40 and 60 min after subcutaneous morphine injection (1 mg/kg). A cut-off was set at 10 s to avoid tissue damage to the rat's tail. The mean latency at peak effect (20 min) was used for the determination of the analgesic efficacy and was converted to % of maximal possible effect (MPE) according to the formula: % maximum possible effect = [(test latency) − (baseline latency)]/[(cut-off) − (baseline latency)] × 100.

*2.5. Molecular modelling*

All in silico studies were performed using Molecular Operating Environment (MOE), MOE2022.02 (Molecular Operating Environment (MOE), 2022.02, Chemical Computing Group ULC, 910-1010 Sherbrooke St. W., Montreal, QC H3A 2R7, 2024)

*2.5.1. Preparation of the receptors*

The 3D coordinates of the β-arrestin 1 and β-arrestin 2 active structures were obtained from Protein Data Bank PDB IDs 4JQI and 8I10, respectively, and loaded into MOE. The missing amino acids were built based on the sequences provided in the Uniprot database (βarr1: P49407, βarr2: P32121). Polar hydrogens and partial charges were added. For the protonation process, a temperature of 300 K, a salt concentration in the solvent of 0.1 mol/L, and pH= 7 were specified. The missing atoms, alternate geometry, and other crystallographic artifacts were corrected by performing QuickPrep. Then, the structure was energy-minimized in the Amber10:EHT force field to an RMS gradient of 0.1 kcal/mol.

*2.5.2. Preparation of the V2Rpp and ARIP peptide*

Each phosphoserine or phosphothreonine of V2Rpp was mutated to a glutamic acid using MOE Protein Builder tool. The structure was protonated, and the partial charges were calculated to assign ionization states and position hydrogens in a macromolecular structure given its 3D coordinates. The energy was minimized to a root mean square (RMS) gradient of 0.1 kcal/mol and Amber10:EHT force field.

*2.6. Statistical analyses*

All data obtained for this study were plotted onto graphs using GraphPad Prism 9 software (RRID:SCR_002798) and represent the mean ± SEM of at least three independent experiments. Sigmoidal concentration-response curves were plotted using the "log(agonist) vs. response (three-parameters)" regression. $pEC_{50}$ and $E_{max}$ values were compared using one-way ANOVA tests with Dunnett's correction for multiple comparisons. Concentration-response curves were compared with respect to ARIP concentration for a given receptor/β-arrestin couple using two-way ANOVA tests. For *in vivo* behavioural tests, the group size was determined as $n = 7$. Animals were randomly assigned to vehicle or to ARIP-treated groups by block randomization. Specific *n* values are supplied in figure legends. To compare the mean values between animal groups, two-way ANOVA tests were performed on behavioural data, with post-hoc correction for multiple comparisons. Statistical differences are depicted in the figures by asterisks (*) or sharps (#). One symbol, $p < 0.05$; two symbols, $p < 0.01$; three, $p < 0.001$; four, $p < 0.0001$). Rounding of values presented in Tables follows the approach outlined by E.H. Blackstone.[51]

## 3. Results & Discussion

### *3.1 Design & synthesis of ARIP*

In this study, the need for β-arrestin pharmacological research tools was addressed by designing a non-selective β-arrestin inhibitor that could directly compete with arrestin recruitment at an activated, agonist-bound 7TMR. To do so, we chose the vasopressin type 2 receptor (V2R) as a starting point, since V2R is known to induce strong and stable recruitment of β-arrestins 1 and 2 following stimulation by AVP (thus, a "stable binder").[10] Furthermore, specific phosphorylated residues of the C-terminus of V2R have been identified as critical for β–arrestin binding[52,53] and a crystal structure of β-arrestin 1 in complex with V2Rpp, a synthetic phosphopeptide derived from residues A343 to S371 of the V2R C-terminus, has been resolved.[23] We therefore decided to base the design of our new *A*rrestin *R*ecruitment *I*nhibitory *P*eptide (ARIP) on this specific peptide sequence (see **Table 1**).

Furthermore, as β-arrestins are cytosolic proteins, any competitive β-arrestin inhibitor would need to be localised on the intracellular side of the plasma membrane bilayer to exert its effect. Thus, using a strategy similar to that of pepducins, we chose to conjugate V2Rpp to a palmitic acid at the N-terminus, to allow anchoring in the plasma membrane and translocation to the intracellular leaflet of the membrane via a passive "flip-flop"mechanism. The proposed mode of action for ARIP is illustrated in the **Graphical Abstract,** in which the lipopeptide, once attached to the membrane and turned towards the cytosol, can compete with the activated, agonist-bound 7TMR for β-arrestin binding and thus inhibit β-arrestin recruitment at the 7TMR.

Finally, to overcome the synthesis challenges encountered when using phosphorylated residues (additional protection/deprotection steps, reduced coupling efficiency, possibility of lateral interactions, etc.), we employed the well-characterised strategy of replacing phosphorylable residues with aspartic and/or glutamic acid in order to generate phosphomimetic (or phosphorylated-like) proteins.[54,55] Here, we substituted the p-Ser and p-Thr residues at the 8 phosphosite positions of V2Rpp with glutamic acid (E) residues: 5 (T347), 8 (S350), 15 (S357), 17 (T359), 18 (T360), 20 (S362), 21 (S363), and 22 (S364) (**Table 1**). The resulting lipopeptide, termed ARIP, was synthesised using conventional Fmoc-based solid-phase peptide synthesis (SPPS), as previously reported.[56] See **Supplementary Table S1 and Supplementary Spectra S1 and S2** for further characterisation.

### *3.2 ARIP inhibits β-arrestin 1 and 2 recruitment at select 7TMRs*

To assess the inhibitory activity of ARIP against agonist-induced β-arrestin recruitment, we performed *in vitro* signalling assays based on $BRET^2$ technology, in which we monitored the proximity between donor-tagged 7TMRs (7TMR-RlucII) and acceptor-tagged β-arrestins (βarr1-GFP10, βarr2-GFP10) in transiently transfected HEK293 cells. We selected a panel of five 7TMRs against which to test functional activity of ARIP: vasopressin 2 receptor (V2R), C-X-C chemokine receptor type 4 (CXCR4), apelin receptor (APJ), μ opioid receptor (MOPR), and glucagon-like peptide 1 receptor (GLP1R). These receptors were selected according to the following criteria. Firstly, although our research generally focuses on class A 7TMRs (Rhodopsin family), we sought to include at least one member of another class in our study: hence the inclusion of the class B receptor GLP1R (Secretin family). Secondly, we considered it important that the two typologies defined by Oakley *et al.* ("transient" versus "stable" binders) should be represented in our receptor panel. In this seminal article, V2R and MOPR were identified as stable and transient binders, respectively.[11] Although CXCR4, APJ, and GLP1R have not been formally assigned to either class, studies suggest that GLP1R and APJ are rapidly internalised and recycled, indicative of a transient binding behaviour,[57–59] whereas CXCR4 is processed more slowly within the cell and poorly recycled, which is characteristic of stable binders.[23] Finally, we sought to ensure that the selected 7TMRs could collectively bind the four major G protein subfamilies ($G\alpha_{q/11}$, $G\alpha_s$, $G\alpha_{i/o}$, $G\alpha_{12/13}$). Although V2R is canonically recognised as a $G\alpha_s$-coupled receptor, it has been shown to signal via many other G proteins, including $G\alpha_{i2}$, $G\alpha_z$, $G\alpha_q$, $G\alpha_{12}$, and $G\alpha_{13}$.[60] CXCR4, APJ, and MOPR primarily couple to $G\alpha_{i/o}$ proteins, although CXCR4 has also been linked to $G\alpha_{13}$[31,61] and APJ, to $G\alpha_q$[62,63] and $G\alpha_{12}$.[64] GLP1R has been reported to promote $G\alpha_s$ signalling, and, to a lesser extent, $G\alpha_q$.[65]

In these cell-based assays, we observed that pre-incubation with ARIP significantly reduced the efficacy (and in some cases, potency) with which agonists recruited β-arrestins 1 and 2 to the receptor (**Figure 1, Tables 2 & 3, Supplementary Tables S2-S4**). ARIP was least effective at inhibiting β-arrestin 1 recruitment to V2R, with only 20 ± 11% inhibition measured at 100 μM of ARIP, and no effect on agonist potency. However, its effect on β-arrestin 2 recruitment was significant, with a shift in agonist potency (2.6 ± 0.23-fold increase in $EC_{50}$ values) and a maximal inhibition of 41 ± 3.0%. At the CXCR4 receptor, ARIP was again more effective against β-arrestin 2 than β-arrestin 1, with no significant effect on agonist potency for either arrestin, but with a marked inhibition of 82 ± 5.0% at 25 μM ARIP for β-arrestin 2 compared with 40 ± 12% at 100 μM for β-arrestin 1 recruitment. For the APJ receptor, ARIP caused significant shifts in Apelin-13 $EC_{50}$ values, with 13 ± 6.1- and 9 ± 1.3-fold increases for β-arrestin 1 and β-arrestin 2 with maximum inhibition of 58 ± 6.5% and 56 ± 2.1%, respectively, at 100 μM of ARIP. For MOPR, ARIP induced significant rightward shifts of the DAMGO curves (5 ± 1.6- and 3 ± 1.2-fold), and maximum inhibition values of 49 ± 8.9% and 35 ± 5.7% were observed for β-arrestin 1 and β-arrestin 2 at 50 μM of ARIP. Finally, while ARIP had no effect on GLP-1's potency to recruit β-arrestins to GLP1R, it significantly reduced the $E_{max}$, with maximum inhibition values of 43 ± 6.6 and 40 ± 10% recruitment, respectively.

These data confirm that ARIP can inhibit 7TMR-mediated β-arrestin recruitment as hypothesised, albeit at fairly high concentrations. Barbadin, the only other β-arrestin inhibitor we know of, has a similarly low potency. Indeed, it has been reported to inhibit the β-arrestin 1/AP2 interaction promoted by AVP with an $IC_{50}$ of 19.1 ± 7.6 μM.[21] It should be noted that membrane-tethered allosteric lipopeptides (pepducins) typically present potencies in the micromolar range. Unlike orthosteric ligands with potencies in the nanomolar range, efficacy does not simply depend on the affinity of the peptide for its target. Rather, it reflects the lipopeptide's ability to flip between the outer and inner phospholipid leaflets of the plasma membrane and its 2D mobility in the inner leaflet, which impacts the effective concentration available to interact with cytosolic proteins. There are a few particularly potent pepducins, such as the β2AR-derived pepducin ICL1-9, which acts as a β-arrestin-biased partial allosteric agonist and has a reported $EC_{50}$ of 96 ± 14 nM. However, most pepducins have potencies closer to those of ICL1-4 (1.9 ± 0.5 μM) and ICL1-11 (1.7 ± 0.5 μM),[33] from the same series, or that of ATI-2341, derived from CXCR4 (0.53 ± 0.91 μM).[31]

As indicated above, our *in vitro* results suggest that ARIP's efficacy as a β-arrestin inhibitor varies from receptor to receptor and, presumably, between ligands at the same receptor site. To better understand this behaviour, we aligned the C-terminal domains of the five 7TMRs and observed minimal sequence identity between receptors. The highest degree of identity was 35.3% (V2R vs CXCR4) and the lowest was 10% (MOPR vs GLPR1) (**Figure S1**). This, in conjunction with the different location of serine and threonine residues (i.e. phosphorylation patterns) in these sequences, suggests that ARIP competes for β-arrestin binding with C-tail sequences that may display variable affinities for arrestins. Intriguingly, ARIP was less effective against its cognate receptor V2R. As V2R was chosen on the basis of its high affinity for β-arrestins and the stability of its arrestin/receptor interaction, it may be that this interaction is too strong for the phosphomimetic lipopeptide to compete effectively, at least at these concentrations. In contrast, ARIP was particularly effective against CXCR4 and APJ when considering $E_{max}$ (statistically significant inhibition was observed even at concentrations of 10 μM against βarr2, and a maximum inhibition of 82 ± 5.0% for CXCR4), and against APJ and MOPR when considering the agonists' $EC_{50}$ values (9 and 13-fold greater potency for Apelin-13; 3 and 5-fold increases in DAMGO potency). We are unsure of the reason for the differences observed between β-arrestin 1 and β-arrestin 2, particularly for the "stable binder" receptors. The non-visual arrestins share 78% sequence identity and are highly conserved at the N-domain groove, where V2Rpp binds[23,24]. Although slight differences in affinity for β-arrestins have been reported previously (another feature associated with "transient binders" is a higher affinity for β-arrestin 2 than for β-arrestin 1[11]), we expected ARIP to bind arrestins in the same way, and therefore exert a similar inhibitory effect on them. These varying efficacies could be explained if ARIP is competing with receptors that might themselves have differential affinities for arrestins; consequently, the inhibitory effect of ARIP would also be dichotomous. However, this was not the case for V2R, and we still observed a slightly greater effect against β-arrestin 2. One potential lead may be in the increased flexibility that has been reported for β-arrestin 2 versus β-arrestin 1, despite their sequence similarity.[66]

Perhaps most critically, many aspects of the β-arrestin/receptor interaction and the phosphorylation motifs underlying it remain unresolved. A 2017 study analysing a crystal structure of the rhodopsin/arrestin complex identified and experimentally verified the Px(x)PxxP phosphorylation motif as a common mechanism for phosphorylation-dependent arrestin recruitment.[67] According to this work, V2R possesses two of these phosphorylation codes, while MOPR has only one. CXCR4, in particular, is thought to possess seven codes. A more recent publication comparing multiple solved structures suggests that a PxPP phorphorylation motif may also be a critical factor in β-arrestin binding. For reference, CXCR4 (S339-T342, S344-347) has two PxPP sites in its C-terminal tail, as does the APJ receptor (S335-S338, S345-S348) – the two receptors against which ARIP appears to be most effective. V2R has a single PxPP site in its C-terminal tail (T360-S363), as does GLP1R (S442-S445). MOPR has none, either in its C-terminal end or in its ICL3.[22] Other publications have associated phosphorylation motifs with individual 7TMRs, such as the four serines and threonines within T370 to T379 in the C-tail of MOPR[68], or argue for the importance of multisite phosphorylation motifs. Clearly, deciphering the mechanims of arrestin/receptor interaction (and, thus, how ARIP may compromise it) remains a complex challenge.

### *3.3 ARIP does not recruit β-arrestins by itself and has no effect on G-protein signalling*

Having shown that ARIP inhibits β-arrestin recruitment at multiple 7TMRs, we next investigated the mechanistic details underlying this inhibitory activity, in particular whether direct recruitment of arrestins by ARIP is possible, irrespective of agonist receptor binding. Thus, in **Figure 2**, we present data from $BRET^2$ time-course experiments monitoring β-arrestin 2 recruitment to V2R (**Figure 2a**) and to the plasma membrane (**Figure 2b**) over a 15-min period. In each case, we observed no increase in $BRET^2$ signal (signifying no increase in arrestin recruitment) by ARIP when tested alone at 25 and 100 μM, in contrast to the peptide agonists AVP (10 μM) or Apelin-13 (1 μM), both of which induced strong β-arrestin 2 recruitment. Nevertheless, ARIP inhibited β-arrestin 2 recruitment to the plasma membrane following Apelin-13 stimulation (**Figure 2b**). In addition, we observe a concentration-dependent decrease in β-arrestin 2 recruitment to the APJ receptor following Apelin-13 treatment over the duration of the readout period, with an almost complete return to baseline at 100 μM ARIP (79 ± 2.2% inhibition, when comparing area under the curve (AUC) values) (**Figure 2c**). Therefore, we postulate that ARIP does not directly draw or recruit arrestins from a cytosolic pool; rather, it binds arrestins at the plasma membrane, once migration to the PM has been initiated by 7TMR activation. Although it has been established that 7TMR phosphorylation is a key step in β-arrestin recruitment and binding, the process that drives the arrestins' migration from the cytosol to the plasma membrane remains unclear. Potentially, this exodus could result from G protein-mediated signalling pathways, which ARIP may be unable to promote.

Additionally, we sought to confirm that ARIP's inhibitory action is specific to β-arrestins and does not extend to G protein activities. We considered this to be particularly important, as it has been well documented that pepducins can act as allosteric agonists of 7TMRs to activate G protein signalling pathways.[31,32,56] First, we tested for cAMP production (a downstream indicator of $G\alpha_s$ activation) in HEK293 cells transiently expressing both V2R and the GFP10-EPAC-RlucII construct, whose conformational change induced by cAMP binding results in loss of $BRET^2$. In **Figure 2d,** we observe that treatment with ARIP (100 μM) neither promoted cAMP production, nor inhibited AVP-promoted cAMP production, throughout the time-course experiment. Similarly, in **Figures 2e** and **2f**, ARIP (100 μM) did not appear to activate $G\alpha_i$ proteins ($G\alpha_{i1}$-RlucII, GFP10-Gγ1) in cells expressing CXCR4 and APJ, respectively, nor did it affect CXCL12- and Apelin-13-induced activation of $G\alpha_{i1}$. Taken together, these data confirm that ARIP selectively affects β-arrestin signalling, lending credence to our hypothesised mode of action in the **Graphical Abstract**. Notably, this mode of action differs from that of canonical pepducins, which act as direct allosteric modulators of their target receptor. Here, β-arrestin is the target of ARIP; its modulation of 7TMR signalling would therefore be indirect and competitive. Due to this differing mode of action, we have retained the nomenclature "lipopeptide" to describe ARIP, despite its similar structure to that of pepducins (N-terminal lipidated peptide sequence derived from a 7TMR's intracellular domain). This mode of action also differs significantly from those of currently available pharmacological tools; Barbadin, the only other β-arrestin inhibitor introduced to date, has been reported to inhibit AP-2/β-arrestin interactions, following 7TMR binding. ARIP would therefore be the first agent to act upstream of this process, to reduce the recruitment of β-arrestins to the receptor.

### *3.4 Molecular modelling suggests that ARIP adapts to β-arrestins 1 and 2 in a similar way to V2Rpp*

Molecular modelling studies were then carried out to determine whether ARIP was able to bind β-arrestins in a similar way to V2Rpp, the phosphorylated peptide derived from the V2R C-terminal domain, and whether replacing the phospho-Ser and phospho-Thr residues of V2Rpp with Glu residues had no significant impact on the inhibitory activity of ARIP on β-arrestin recruitment. Two key considerations have guided our initial decision to subsitute Glu for p-Ser and p-Thr. Firstly, we felt that designing a peptide capable of mimicking V2Rpp activity while lacking its phosphorylated residues would greatly faciltate the synthesis process and provide a valuable tool for studying and understanding the interactions between V2Rpp and β-arrestin 1. Secondly, the V2Rpp binding site has been shown to contain basic amino acids that interact with phosphate groups. By introducing acidic amino acids (c.a. glutamic acid) at these positions, we sought to maintain the crucial electrostatic interactions that exist between the phosphorylated residues of V2R and β-arrestin 1. However, we also recognise that these C-terminal p-Ser and p-Thr residues serve as important recognition sites for binding to β-arrestin and form critical contacts with the positively charged Lys and Arg residues in the N-terminal groove of β-arrestin.[23] Conceivably, some of these contacts might be lost by phospho-Ser/Glu and phospho-Thr/Glu replacements, affecting ARIP/βarr binding and, consequently, the efficacy of ARIP as a β-arrestin inhibitor.

To start, we performed molecular docking simulations between V2Rpp, the phosphorylated peptide derived from the C-terminal residues of V2R, and β-arrestin 1. We chose to model the interactions of V2Rpp and ARIP into β-arrestin 1, but not with β-arrestin 2, as these arrestins share high sequence identity (78%) and an alignment of human β-arrestins 1 and 2 revealed that the V2Rpp-binding residues are highly conserved. Indeed, as shown in **Figure 3a**, only 2 residues close to the V2Rpp/βarr binding site differ: in β-arrestin 2, an alanine is mutated to serine at position 12 of β-arrestin 1, while a threonine is mutated to serine, representing a very minor modification, at position 74.

Molecular modelling results revealed that ARIP has the same binding mode as the original V2Rpp in β-arrestin 1 with high complementarity (**Figure 3b**). Consistent with previous work,[23] we found that the positively charged amino groups of the lysine and arginine residues present within the binding site of β-arrestin 1 are strategically positioned to form strong hydrogen bond network and ionic interactions with the negatively charged phosphate groups of phosphoserine and phosphothreonine. The positions that these phosphoserines and phosphothreonines occupy in the C-terminus of V2R (amno acids 343 to 371), as well as the corresponding Glu (E) residues in ARIP (pos. 1-29), are presented in **Figure 3c,** and **Figure 3d** shows the interactions between each phosphorylated amino acid in V2Rpp with β-arrestin 1 separately. Our modelling data highlights the importance of phosphate groups in specific positions, notably positions 15, 18, 21, and 22, to form key contacts with β-arrestin 1, which is consistent with the literature.[25]

To investigate the impact of Glu substitutions, we conducted a molecular docking study between our lipidated phosphomimetic peptide, ARIP, and β-arrestin 1, comparing key interactions with those of V2Rpp/β-arrestin 1. Comparing Glu347 to Thr($PO_3H_2$)347, we observe that Glu347 maintains two hydrogen bonds with Lys77 and Arg65 (as shown in **Figure 3e, '5'**). However, it appears that Glu347 is not positioned close enough to interact with Tyr63. For interactions with Glu350, amino acids Arg62, Arg165, Thr74, and Lys138 form a hydrogen bond (**Figure 3e, '8'**). Compared with Ser($PO_3H_2$)350, it is apparent that the molecule's conformation is slightly different. In this case, instead of Arg65, Glu350 forms a hydrogen bond with Arg62. Nevertheless, the overall interactions are preserved. Similarly, the mutation of Ser($PO_3H_2$)357 to Glu357 does not seem to affect the interaction with Lys11, Lys138, Lys160, and Arg165 (**Figure 3e, '15'**). Lys10 is the only basic amino acid involved in the interaction with Glu359, as is the case with Thr($PO_3H_2$)359 (**Figure 3e, '17'**). Additionally, Lys11, Lys294, and Arg25 form hydrogen interactions with Glu360 (**Figure 3e, '18'**). On the other hand, comparing Glu362 and 363 to Ser($PO_3H_2$)362 and 363, Glu362 and 363 do not establish hydrogen bonds with Lys10 (**Figure 3e, '20'** and **'21'**). Finally, Glu364 exhibits a consistent interaction with Lys107 compared to Ser($PO_3H_2$)364 (**Figure 3e, '22'**).

Overall, in comparison to V2Rpp, we observed that some hydrogen bond interactions between ARIP and β-arrestin 1 are hindered by the absence of the phosphoryl groups, in particular the mutations at positions **18** (T360E), **20** (S362E), and **21** (S363E). However, it is noteworthy that in the case of ARIP, replacement of phosphorylated residues with glutamic acid demonstrated that these substitutions are sufficient to allow ARIP to potentially bind to the same binding site as V2Rpp. This suggests that the introduction of acidic amino acids can mimic the electrostatic interactions typically mediated by phosphorylated residues. This is consistent with previous reports, in particular a study by Dwivedi-Agnihotri *et al.* in 2020,[25] in which the authors introduced a series of alanine mutations into V2R's C-terminal phosphosites that negatively impacted β-arrestin recruitment in mutants S357A (15), T360A (18), S362A (20) and S363A (21). In light of this study and of our own modelling data, we chose to synthesise a partially phosphorylated version of ARIP (ARIP-4P), in which the appropriate p-Ser and p-Thr residues were inserted in positions **15** (S357), **18** (T360), **20** (S362) and **21** (S363), and the phosphomimetic Glu (E) residues were conserved in the remaining sites (**Figure 4a**).

### *3.5 ARIP and ARIP-4P share similar inhibitory activities against β-arrestin recruitment to 7TMRs*

Like ARIP, ARIP-4P was synthesised on solid phase using an automated peptide synthesiser and a similar coupling procedure, with an alternate deprotecting solution to prevent the formation of β-elimination by-products (for characterisation of ARIP-4P, see **Supplementary Table S1** and **Supplementary Spectra S3** and **S4**). The resulting compound was assessed in the same *in vitro* β-arrestin recruitment assays as those performed with ARIP **(Figure 4a**).

We hypothesised that ARIP-4P might have an increased affinity for β-arrestins and might therefore exert greater β-arrestin inhibition than ARIP in $BRET^2$ assays. As shown in **Figure 4 & Table 4**, the inhibitory effect of ARIP on β-arrestin 2 recruitment to APJ after stimulation by Apelin-13 was very similar to that of ARIP, if not slightly less effective. Indeed, ARIP-4P significantly reduced Apelin-13-promoted β-arrestin 2 recruitment at concentrations 25 μM and above, with a max inhibition of 57 ± 4.5% (**Figure 4b**). In the same experimental paradigm, ARIP reached a max inhibition 81 ± 9.1% at 100 μM (**Supplementary Table S5**). Like ARIP, ARIP-4P alone did not recruit β-arrestins to the plasma membrane but was effective in inhibiting β-arrestin 2 recruitment to the plasma membrane following stimulation by Apelin-13 (**Figure 4c**).

Altogether, our decision to substitute the p-Ser and p-Thr of V2Rpp by Glu residues when designing ARIP was found to be largely effective, facilitating the process of lipopeptide synthesis while retaining many of the critical contacts modelled between V2Rpp and β-arrestins. Alhough our docking study suggests that greater affinity between ARIP and β-arrestins could be achieved by preserving the initial phosphorylated residues in key positions, our results with ARIP-4P did not reveal any increase in β-arrestin inhibition. One possible explanation is that the addition of bulky and heavily-charged phosphate groups, although advantageous for ARIP-4P/arrestin binding, is counterbalanced by reduced permeability at the cell membrane, a less efficient "flip-flop" process that is a critical first step in ARIP's mode of action. In addition, by its very design, ARIP-4P is vulnerable to the action of cytosolic phosphatases, which is not the case for ARIP, with its Glu substitutions. Thus, the ARIP-4P lipopeptide, once having crossed the cell membrane barrier, may be rapidly dephosphoryated and lose its efficacy.

### *3.6 ARIP potentiates the analgesic effect of morphine after intrathecal delivery through β-arrestin inhibition*

As ARIP effectively inhibited the recruitment of β-arrestin to each of the selected 7TMRs in our *in vitro* experiments, we next sought to study whether this result could be translated to an *in vivo* setting. Our goal was therefore to produce an ARIP-mediated behavioural response that would be consistent with β-arrestin inhibition, in an animal model. However, to do so, the link between β-arrestin inhibition and a particular physiological response must first have been established; for many of the 7TMRs in our panel, the contribution of β-arrestin signalling to their various physiological effects has yet to be determined. In the case of MOPR, previous work by Bohn and colleagues[36] demonstrated that functional deletion of the gene coding for β–arrestin 2 (βarr2$^{-/-}$) in mice potentiated and prolonged the analgesic effect of morphine in a model of acute thermal pain. This potentiation was observed at both analgesic and subanalgesic doses of morphine. The authors theorised that this was due to reduced receptor desensitisation: in βarr2$^{-/-}$ mice, arrestin-mediated receptor endocytosis would be impaired, resulting in increased retention of MOPR at the cell membrane and, consequently, more potent and longer-lasting morphine-induced analgesia.

To reproduce this effect pharmacologically, ARIP was injected intrathecally into Sprague-Dawley rats under a similar acute thermal pain paradigm (tail-flick test) (**Figure 5**). Here, the rat's latency to withdraw its tail from a nociceptive stimulus (a hot water bath maintained at 52$^{o}$C) was assessed 12 hours post i.t. injection of ARIP (250 nmol/kg) or its vehicle and for 1 hour after s.c. injection of morphine (1 mg/kg) or saline. The relatively low dose of morphine (1 mg/kg) was specifically chosen in order to produce a mild analgesia to ensure that the potentiation by ARIP would be evident. As shown in **Figure 5a,** pre-administration of ARIP did not significantly alter the baseline tail withdrawal latencies between the "Vehicle + Saline" and "ARIP + Saline" groups. However, we observed a potent and sustained increase in tail withdrawal latencies between the "Vehicle + Morphine" and "ARIP + Morphine" groups, representing a 103 ± 8 % increase when comparing total area under the curve (AUC) values, or a 72 ± 8 % increase when comparing the maximal possible effect (MPE) values, 20 minutes after morphine injection (**Figure 5b**). Taken together, these data demonstrate that i.t. pre-treatment with ARIP enhanced and prolonged the antinociceptive effect of morphine, for at least 1 hour post-injection. These results are consistent with those obtained by Bohn et al[36] using β-arrestin 2 knockout mice, reinforcing the role of ARIP effectiveness as a β-arrestin inhibitor and providing important *in vivo* proof-of-concept for the present study.

## 4. Conclusion

In summary, we have designed, synthesised, and characterised a novel phosphomimetic lipopeptide inhibitor derived from the C-terminal domain of V2R, named ARIP. In cell-based assays monitoring the recruitment of β-arrestins to 7TMRs, we observed an inhibitory effect at all the receptors tested (V2R, CXCR4, APJ, MOPR, and GLP1R), with varying degrees of potency and efficacy. We also demonstrated that ARIP does not recruit β-arrestins to the plasma membrane itself, nor does it affect canonical G protein signalling. These data suggest that ARIP does not act as a direct allosteric agonist or allosteric modulator of these 7TMRs, unlike similar lipidated peptides, such as pepducins.[27,28] In addition, molecular modelling data revealed that ARIP binds with high complementarity to β-arrestins akin to V2Rpp and that substitution of p-Ser and p-Thr residues 15, 18, 20, and 21 of V2Rpp by Glu residues had no significant impact on the β-arrestin inhibitory effectiveness of ARIP. Importantly, we provided evidence of ARIP activity *in vivo*, with spinal injection of ARIP potentiating the analgesic action of morphine in the acute thermal pain test, a behavioural response compatible with β-arrestin inhibition. Although further characterisation of ARIP is advisable, for example, the ability of ARIP to activate arrestin-specific downstream signalling pathways, to inhibit 7TMR endocytosis, or to interact with receptors whose arrestin recruitment mechanisms are phosphorylation-independent (i.e., M1R[69], SPR[70] or BLTR1[71]), remains to be assessed, we believe that ARIP holds great promise as a new inhibitor of β-arrestin recruitment. Notably, it would be the first to directly inhibit recruitment at the 7TMR, in contrast to Barbadin, which is believed to inhibit AP-2/β-arrestin interactions. Thus, ARIP may serve as an important pharmacological tool, helping to decipher the relative contributions of β-arrestin signalling in 7TMR physiology and pathophysiology, thereby potentially contributing to the development of better therapeutics.

**Acknowledgements**

The authors would like to recognise the valuable contributions of our late colleague, Pr Éric Marsault (Department of Pharmacology-Physiology, Université de Sherbrooke), to this project. We would like to thank Prs. M. Bouvier, T. Hebert, S.A. Laporte, G. Pineyro, R. Leduc, J.-C. Tardif and E. Thorin for providing us with $BRET^2$-based biosensors. As well, we would like to thank Drs. N. Heveker, L. Gendron, and R. Jorgensen for the CXCR4-RlucII, MOPR-RlucII and GLP1R-RlucII plasmids, respectively. R.L.B. is the recipient of Ph.D. scholarships from the *Fonds de recherche du Québec – Santé (FRQ-S)* and the Canadian Institutes of Health Research (CIHR). M.D. is the recipient of a Ph.D. scholarship from the FRQ-S. É.B. is a recipient of Master's-level scholarships from the FRQ-S and the CIHR, and F.L, from the CIHR. P.-L.B. is a member of the FRQ-NT-funded PROTEO network. E.B.-O. is the recipient of an Excellence Research Chair in Innovative Theranostic Approaches in Ovarian Cancers (THERANOVCA) funded by the Région Normandie and cofinanced by the European Union. P.S. is the recipient of a Tier 1 Canada Research Chair in Neurophysiopharmacology of Chronic Pain and a member of the FRQ-S-funded Québec Pain Research Network (QPRN).

**Funding sources**

This research was funded by the Canadian Institutes of Health Research (CIHR), grant number FDN-148413.

**Conflicts of interest**

The authors declare no competing financial interests.

**Author contributions**

Conceptualization, C.E.M. and E.B.-O.; Methodology, R.L.B., C.E.M., M.D., M.H., P.-L.B., P.S., and E.B.-O.; Validation, R.L.B., C.E.M., J.-M.L., M.G., P.-L.B., P.S., and E.B.-O.; Formal analysis, R.L.B., C.E.M., E.B.-O., and P.-L.B.; Investigation, R.L.B., M.D., E.B., F.L, M.H., K.B., and E.B.-O.; Resources, C.E.M. and E.B.-O.; Data curation, R.L.B., M.H, J.-M.L.; Writing—original draft preparation, R.L.B.; Writing—review and editing, R.L.B., C.E.M., J.-M.L., M.G., P.-L.B., P.S., and E.B.-O.; Visualization, R.L.B. and E.B.-O.; Supervision, C.E.M., P.S., and E.B.-O.; Project administration, C.E.M., P.S., and E.B.-O.; Funding acquisition, P.S. All authors have read and agreed to the published version of the manuscript.

## References

[1] R. Sterne-Marr, V.V. Gurevich, P. Goldsmith, R.C. Bodine, C. Sanders, L.A. Donoso, J.L. Benovic, Polypeptide variants of beta-arrestin and arrestin3, J. Biol. Chem. 268 (1993) 15640–15648.

[2] H. Attramadal, J.L. Arriza, C. Aoki, T.M. Dawson, J. Codina, M.M. Kwatra, S.H. Snyder, M.G. Caron, R.J. Lefkowitz, Beta-arrestin2, a novel member of the arrestin/beta-arrestin gene family, J. Biol. Chem. 267 (1992) 17882–17890.

[3] M.J. Lohse, J.L. Benovic, J. Codina, M.G. Caron, R.J. Lefkowitz, β-Arrestin: a Protein that Regulates β-adrenergic Receptor Function, Science 248 (1990) 1547–1550. https://doi.org/10.1126/science.2163110.

[4] V.V. Gurevich, E.V. Gurevich, GPCR Signaling Regulation: The Role of GRKs and Arrestins, Front. Pharmacol. 10 (2019) 125. https://doi.org/10.3389/fphar.2019.00125.

[5] K.L. Pierce, R.J. Lefkowitz, Classical and new roles of β-arrestins in the regulation of G-PROTEIN-COUPLED receptors, Nat. Rev. Neurosci. 2 (2001) 727–733. https://doi.org/10.1038/35094577.

[6] A.S. Hauser, M.M. Attwood, M. Rask-Andersen, H.B. Schiöth, D.E. Gloriam, Trends in GPCR drug discovery: new agents, targets and indications, Nat. Rev. Drug Discov. 16 (2017) 829–842. https://doi.org/10.1038/nrd.2017.178.

[7] R. Santos, O. Ursu, A. Gaulton, A.P. Bento, R.S. Donadi, C.G. Bologa, A. Karlsson, B. Al-Lazikani, A. Hersey, T.I. Oprea, J.P. Overington, A comprehensive map of molecular drug targets, Nat. Rev. Drug Discov. 16 (2017) 19–34. https://doi.org/10.1038/nrd.2016.230.

[8] K. Sriram, P.A. Insel, G Protein-Coupled Receptors as Targets for Approved Drugs: How Many Targets and How Many Drugs?, Mol. Pharmacol. 93 (2018) 251–258. https://doi.org/10.1124/mol.117.111062.

[9] L.M. Luttrell, F.L. Roudabush, E.W. Choy, W.E. Miller, M.E. Field, K.L. Pierce, R.J. Lefkowitz, Activation and targeting of extracellular signal-regulated kinases by β-arrestin scaffolds, Proc. Natl. Acad. Sci. 98 (2001) 2449–2454. https://doi.org/10.1073/pnas.041604898.

[10] R.H. Oakley, S.A. Laporte, J.A. Holt, L.S. Barak, M.G. Caron, Association of β-Arrestin with G Protein-coupled Receptors during Clathrin-mediated Endocytosis Dictates the Profile of Receptor Resensitization, J. Biol. Chem. 274 (1999) 32248–32257. https://doi.org/10.1074/jbc.274.45.32248.

[11] R.H. Oakley, S.A. Laporte, J.A. Holt, M.G. Caron, L.S. Barak, Differential Affinities of Visual Arrestin, βArrestin1, and βArrestin2 for G Protein-coupled Receptors Delineate Two Major Classes of Receptors, J. Biol. Chem. 275 (2000) 17201–17210. https://doi.org/10.1074/jbc.M910348199.

[12] R.H. Oakley, S.A. Laporte, J.A. Holt, L.S. Barak, M.G. Caron, Molecular Determinants Underlying the Formation of Stable Intracellular G Protein-coupled Receptor-β-Arrestin Complexes after Receptor Endocytosis*, J. Biol. Chem. 276 (2001) 19452–19460. https://doi.org/10.1074/jbc.M101450200.

[13] T.J. Cahill, A.R.B. Thomsen, J.T. Tarrasch, B. Plouffe, A.H. Nguyen, F. Yang, L.-Y. Huang, A.W. Kahsai, D.L. Bassoni, B.J. Gavino, J.E. Lamerdin, S. Triest, A.K. Shukla, B. Berger, J. Little, A. Antar, A. Blanc, C.-X. Qu, X. Chen, K. Kawakami, A. Inoue, J. Aoki, J. Steyaert, J.-P. Sun, M. Bouvier, G. Skiniotis, R.J. Lefkowitz, Distinct conformations of

GPCR–β-arrestin complexes mediate desensitization, signaling, and endocytosis, Proc. Natl. Acad. Sci. 114 (2017) 2562–2567. https://doi.org/10.1073/pnas.1701529114.

[14] N.R. Latorraca, M. Masureel, S.A. Hollingsworth, F.M. Heydenreich, C.-M. Suomivuori, C. Brinton, R.J.L. Townshend, M. Bouvier, B.K. Kobilka, R.O. Dror, How GPCR Phosphorylation Patterns Orchestrate Arrestin-Mediated Signaling, Cell 183 (2020) 1813-1825.e18. https://doi.org/10.1016/j.cell.2020.11.014.

[15] S. Mangmool, H. Kurose, G(i/o) protein-dependent and -independent actions of Pertussis Toxin (PTX), Toxins 3 (2011) 884–899. https://doi.org/10.3390/toxins3070884.

[16] V. Inamdar, A. Patel, B.K. Manne, C. Dangelmaier, S.P. Kunapuli, Characterization of UBO-QIC as a Gαq inhibitor in platelets, Platelets 26 (2015) 771–778. https://doi.org/10.3109/09537104.2014.998993.

[17] Y. Zhu, H. Chen, S. Boulton, F. Mei, N. Ye, G. Melacini, J. Zhou, X. Cheng, Biochemical and Pharmacological Characterizations of ESI-09 Based EPAC Inhibitors: Defining the ESI-09 "Therapeutic Window," Sci. Rep. 5 (2015) 9344. https://doi.org/10.1038/srep09344.

[18] S. Narumiya, T. Ishizaki, M. Uehata, Use and properties of ROCK-specific inhibitor Y-27632, Methods Enzymol. 325 (2000) 273–284. https://doi.org/10.1016/s0076-6879(00)25449-9.

[19] E. Macia, M. Ehrlich, R. Massol, E. Boucrot, C. Brunner, T. Kirchhausen, Dynasore, a Cell-Permeable Inhibitor of Dynamin, Dev. Cell 10 (2006) 839–850. https://doi.org/10.1016/j.devcel.2006.04.002.

[20] T.A. Hill, C.P. Gordon, A.B. McGeachie, B. Venn-Brown, L.R. Odell, N. Chau, A. Quan, A. Mariana, J.A. Sakoff, M. Chircop (nee Fabbro), P.J. Robinson, A. McCluskey, Inhibition of Dynamin Mediated Endocytosis by the *Dynoles* —Synthesis and Functional Activity of a

Family of Indoles, J. Med. Chem. 52 (2009) 3762–3773.
https://doi.org/10.1021/jm900036m.

[21] A. Beautrait, J.S. Paradis, B. Zimmerman, J. Giubilaro, L. Nikolajev, S. Armando, H. Kobayashi, L. Yamani, Y. Namkung, F.M. Heydenreich, E. Khoury, M. Audet, P.P. Roux, D.B. Veprintsev, S.A. Laporte, M. Bouvier, A new inhibitor of the β-arrestin/AP2 endocytic complex reveals interplay between GPCR internalization and signalling, Nat. Commun. 8 (2017) 15054. https://doi.org/10.1038/ncomms15054.

[22] J. Maharana, P. Sarma, M.K. Yadav, S. Saha, V. Singh, S. Saha, M. Chami, R. Banerjee, A.K. Shukla, Structural snapshots uncover a key phosphorylation motif in GPCRs driving β-arrestin activation, Mol. Cell (2023) S1097276523003209.
https://doi.org/10.1016/j.molcel.2023.04.025.

[23] A.K. Shukla, A. Manglik, A.C. Kruse, K. Xiao, R.I. Reis, W.-C. Tseng, D.P. Staus, D. Hilger, S. Uysal, L.-Y. Huang, M. Paduch, P. Tripathi-Shukla, A. Koide, S. Koide, W.I. Weis, A.A. Kossiakoff, B.K. Kobilka, R.J. Lefkowitz, Structure of active β-arrestin-1 bound to a G-protein-coupled receptor phosphopeptide, Nature 497 (2013) 137–141.
https://doi.org/10.1038/nature12120.

[24] Q.-T. He, P. Xiao, S.-M. Huang, Y.-L. Jia, Z.-L. Zhu, J.-Y. Lin, F. Yang, X.-N. Tao, R.-J. Zhao, F.-Y. Gao, X.-G. Niu, K.-H. Xiao, J. Wang, C. Jin, J.-P. Sun, X. Yu, Structural studies of phosphorylation-dependent interactions between the V2R receptor and arrestin-2, Nat. Commun. 12 (2021) 2396. https://doi.org/10.1038/s41467-021-22731-x.

[25] H. Dwivedi-Agnihotri, M. Chaturvedi, M. Baidya, T.M. Stepniewski, S. Pandey, J. Maharana, A. Srivastava, N. Caengprasath, A.C. Hanyaloglu, J. Selent, A.K. Shukla, Distinct phosphorylation sites in a prototypical GPCR differently orchestrate β-arrestin

interaction, trafficking, and signaling, Sci. Adv. 6 (2020) eabb8368. https://doi.org/10.1126/sciadv.abb8368.

[26] L. Covic, A.L. Gresser, J. Talavera, S. Swift, A. Kuliopulos, Activation and inhibition of G protein-coupled receptors by cell-penetrating membrane-tethered peptides, Proc. Natl. Acad. Sci. 99 (2002) 643–648. https://doi.org/10.1073/pnas.022460899.

[27] K.E. Carlson, T.J. McMurry, S.W. Hunt, Pepducins: lipopeptide allosteric modulators of GPCR signaling, Drug Discov. Today Technol. 9 (2012) e33–e39. https://doi.org/10.1016/j.ddtec.2011.07.002.

[28] R. Carr, J.L. Benovic, From biased signalling to polypharmacology: unlocking unique intracellular signalling using pepducins, Biochem. Soc. Trans. 44 (2016) 555–561. https://doi.org/10.1042/BST20150230.

[29] S.J.H. Wielders, A. Bennaghmouch, C.P.M. Reutelingsperger, E.M. Bevers, T. Lindhout, Anticoagulant and antithrombotic properties of intracellular protease-activated receptor antagonists, J. Thromb. Haemost. 5 (2007) 571–576. https://doi.org/10.1111/j.1538-7836.2007.02364.x.

[30] M. Tsuji, S. Ueda, T. Hirayama, K. Okuda, Y. Sakaguchi, A. Isono, H. Nagasawa, FRET-based imaging of transbilayer movement of pepducin in living cells by novel intracellular bioreductively activatable fluorescent probes, Org. Biomol. Chem. 11 (2013) 3030. https://doi.org/10.1039/c3ob27445d.

[31] J. Quoyer, J.M. Janz, J. Luo, Y. Ren, S. Armando, V. Lukashova, J.L. Benovic, K.E. Carlson, S.W. Hunt, M. Bouvier, Pepducin targeting the C-X-C chemokine receptor type 4 acts as a biased agonist favoring activation of the inhibitory G protein, Proc. Natl. Acad. Sci. 110 (2013) E5088–E5097. https://doi.org/10.1073/pnas.1312515110.

[32] R. Carr, Y. Du, J. Quoyer, R.A. Panettieri, J.M. Janz, M. Bouvier, B.K. Kobilka, J.L. Benovic, Development and Characterization of Pepducins as $G_s$ -biased Allosteric Agonists, J. Biol. Chem. 289 (2014) 35668–35684. https://doi.org/10.1074/jbc.M114.618819.

[33] R. Carr, J. Schilling, J. Song, R.L. Carter, Y. Du, S.M. Yoo, C.J. Traynham, W.J. Koch, J.Y. Cheung, D.G. Tilley, J.L. Benovic, β-arrestin–biased signaling through the $\beta_2$ -adrenergic receptor promotes cardiomyocyte contraction, Proc. Natl. Acad. Sci. 113 (2016) E4107–E4116. https://doi.org/10.1073/pnas.1606267113.

[34] H. Nassour, T.A. Hoang, R.D. Martin, J.C.C. Dallagnol, É. Billard, M. Létourneau, E. Novellino, A. Carotenuto, B.G. Allen, J.C. Tanny, A. Fournier, T.E. Hébert, D. Chatenet, Lipidated peptides derived from intracellular loops 2 and 3 of the urotensin II receptor act as biased allosteric ligands, J. Biol. Chem. 297 (2021) 101057. https://doi.org/10.1016/j.jbc.2021.101057.

[35] L.A. Grisanti, T.P. Thomas, R.L. Carter, C. de Lucia, E. Gao, W.J. Koch, J.L. Benovic, D.G. Tilley, Pepducin-mediated cardioprotection via β-arrestin-biased β2-adrenergic receptor-specific signaling, Theranostics 8 (2018) 4664–4678. https://doi.org/10.7150/thno.26619.

[36] L.M. Bohn, R.J. Lefkowitz, R.R. Gainetdinov, K. Peppel, M.G. Caron, F.-T. Lin, Enhanced Morphine Analgesia in Mice Lacking β-Arrestin 2, Science 286 (1999) 2495–2498. https://doi.org/10.1126/science.286.5449.2495.

[37] T.J. Attard, N.M. O'Brien-Simpson, E.C. Reynolds, Identification and Suppression of β-Elimination Byproducts Arising from the Use of Fmoc-Ser(PO3Bzl,H)-OH in Peptide Synthesis, Int. J. Pept. Res. Ther. 15 (2009) 69–79. https://doi.org/10.1007/s10989-008-9165-9.

[38] C. Ehrhardt, M. Schmolke, A. Matzke, A. Knoblauch, C. Will, V. Wixler, S. Ludwig, Polyethylenimine, a cost-effective transfection reagent, Signal Transduct. 6 (2006) 179–184. https://doi.org/10.1002/sita.200500073.

[39] R. Jorgensen, S. Norklit Roed, A. Heding, C.E. Elling, Beta-arrestin2 as a competitor for GRK2 interaction with the GLP-1 receptor upon receptor activation, Pharmacology 88 (2011) 174–181. https://doi.org/10.1159/000330742.

[40] C.E. Mona, É. Besserer-Offroy, J. Cabana, R. Leduc, P. Lavigne, N. Heveker, É. Marsault, E. Escher, Design, synthesis, and biological evaluation of CXCR4 ligands, Org. Biomol. Chem. 14 (2016) 10298–10311. https://doi.org/10.1039/C6OB01484D.

[41] J.-L. Beaudeau, V. Blais, B.J. Holleran, A. Bergeron, G. Piñeyro, B. Guérin, L. Gendron, Y.L. Dory, *N* -Guanidyl and *C* -Tetrazole Leu-Enkephalin Derivatives: Efficient Mu and Delta Opioid Receptor Agonists with Improved Pharmacological Properties, ACS Chem. Neurosci. 10 (2019) 1615–1626. https://doi.org/10.1021/acschemneuro.8b00550.

[42] Y. Namkung, C. Le Gouill, V. Lukashova, H. Kobayashi, M. Hogue, E. Khoury, M. Song, M. Bouvier, S.A. Laporte, Monitoring G protein-coupled receptor and β-arrestin trafficking in live cells using enhanced bystander BRET, Nat. Commun. 7 (2016) 12178. https://doi.org/10.1038/ncomms12178.

[43] B. Zimmerman, A. Beautrait, B. Aguila, R. Charles, E. Escher, A. Claing, M. Bouvier, S.A. Laporte, Differential β-arrestin-dependent conformational signaling and cellular responses revealed by angiotensin analogs, Sci. Signal. 5 (2012) ra33. https://doi.org/10.1126/scisignal.2002522.

[44] J.S. Paradis, S. Ly, É. Blondel-Tepaz, J.A. Galan, A. Beautrait, M.G.H. Scott, H. Enslen, S. Marullo, P.P. Roux, M. Bouvier, Receptor sequestration in response to β-arrestin-2 phosphorylation by ERK1/2 governs steady-state levels of GPCR cell-surface expression,

Proc. Natl. Acad. Sci. U. S. A. 112 (2015) E5160-5168. https://doi.org/10.1073/pnas.1508836112.

[45] B. Breton, É. Sauvageau, J. Zhou, H. Bonin, C. Le Gouill, M. Bouvier, Multiplexing of Multicolor Bioluminescence Resonance Energy Transfer, Biophys. J. 99 (2010) 4037–4046. https://doi.org/10.1016/j.bpj.2010.10.025.

[46] C. Galés, R.V. Rebois, M. Hogue, P. Trieu, A. Breit, T.E. Hébert, M. Bouvier, Real-time monitoring of receptor and G-protein interactions in living cells, Nat. Methods 2 (2005) 177–184. https://doi.org/10.1038/nmeth743.

[47] S. Armando, J. Quoyer, V. Lukashova, A. Maiga, Y. Percherancier, N. Heveker, J. Pin, L. Prézeau, M. Bouvier, The chemokine CXC4 and CC2 receptors form homo- and heterooligomers that can engage their signaling G-protein effectors and βarrestin, FASEB J. 28 (2014) 4509–4523. https://doi.org/10.1096/fj.13-242446.

[48] M. Leduc, B. Breton, C. Galés, C. Le Gouill, M. Bouvier, S. Chemtob, N. Heveker, Functional Selectivity of Natural and Synthetic Prostaglandin $EP_4$ Receptor Ligands, J. Pharmacol. Exp. Ther. 331 (2009) 297–307. https://doi.org/10.1124/jpet.109.156398.

[49] É. Besserer-Offroy, P. Bérubé, J. Côté, A. Murza, J.-M. Longpré, R. Dumaine, O. Lesur, M. Auger-Messier, R. Leduc, É. Marsault, P. Sarret, The hypotensive effect of activated apelin receptor is correlated with β-arrestin recruitment, Pharmacol. Res. 131 (2018) 7–16. https://doi.org/10.1016/j.phrs.2018.02.032.

[50] N. Percie du Sert, V. Hurst, A. Ahluwalia, S. Alam, M.T. Avey, M. Baker, W.J. Browne, A. Clark, I.C. Cuthill, U. Dirnagl, M. Emerson, P. Garner, S.T. Holgate, D.W. Howells, N.A. Karp, S.E. Lazic, K. Lidster, C.J. MacCallum, M. Macleod, E.J. Pearl, O.H. Petersen, F. Rawle, P. Reynolds, K. Rooney, E.S. Sena, S.D. Silberberg, T. Steckler, H. Würbel, The

ARRIVE guidelines 2.0: updated guidelines for reporting animal research, J. Physiol. 598 (2020) 3793–3801. https://doi.org/10.1113/JP280389.

[51] E.H. Blackstone, Rounding numbers, J. Thorac. Cardiovasc. Surg. 152 (2016) 1481–1483. https://doi.org/10.1016/j.jtcvs.2016.09.003.

[52] K. Xiao, S.K. Shenoy, K. Nobles, R.J. Lefkowitz, Activation-dependent Conformational Changes in β-Arrestin 2, J. Biol. Chem. 279 (2004) 55744–55753. https://doi.org/10.1074/jbc.M409785200.

[53] K.N. Nobles, Z. Guan, K. Xiao, T.G. Oas, R.J. Lefkowitz, The Active Conformation of β-Arrestin1, J. Biol. Chem. 282 (2007) 21370–21381. https://doi.org/10.1074/jbc.M611483200.

[54] W. Huang, R.L. Erikson, Constitutive activation of Mek1 by mutation of serine phosphorylation sites., Proc. Natl. Acad. Sci. 91 (1994) 8960–8963. https://doi.org/10.1073/pnas.91.19.8960.

[55] N.M. Otto, W.G. McDowell, D.M. Dickey, L.R. Potter, A Glutamate-Substituted Mutant Mimics the Phosphorylated and Active Form of Guanylyl Cyclase-A, Mol. Pharmacol. 92 (2017) 67–74. https://doi.org/10.1124/mol.116.107995.

[56] R.L. Brouillette, É. Besserer-Offroy, C.E. Mona, M. Chartier, S. Lavenus, M. Sousbie, K. Belleville, J.-M. Longpré, É. Marsault, M. Grandbois, P. Sarret, Cell-penetrating pepducins targeting the neurotensin receptor type 1 relieve pain, Pharmacol. Res. 155 (2020) 104750. https://doi.org/10.1016/j.phrs.2020.104750.

[57] D.K. Lee, S.S.G. Ferguson, S.R. George, B.F. O'Dowd, The fate of the internalized apelin receptor is determined by different isoforms of apelin mediating differential interaction with β-arrestin, Biochem. Biophys. Res. Commun. 395 (2010) 185–189. https://doi.org/10.1016/j.bbrc.2010.03.151.

[58] D. Bhandari, J. Trejo, J.L. Benovic, A. Marchese, Arrestin-2 Interacts with the Ubiquitin-Protein Isopeptide Ligase Atrophin-interacting Protein 4 and Mediates Endosomal Sorting of the Chemokine Receptor CXCR4, J. Biol. Chem. 282 (2007) 36971–36979. https://doi.org/10.1074/jbc.M705085200.

[59] S.N. Roed, P. Wismann, C.R. Underwood, N. Kulahin, H. Iversen, K.A. Cappelen, L. Schäffer, J. Lehtonen, J. Hecksher-Soerensen, A. Secher, J.M. Mathiesen, H. Bräuner-Osborne, J.L. Whistler, S.M. Knudsen, M. Waldhoer, Real-time trafficking and signaling of the glucagon-like peptide-1 receptor, Mol. Cell. Endocrinol. 382 (2014) 938–949. https://doi.org/10.1016/j.mce.2013.11.010.

[60] F.M. Heydenreich, B. Plouffe, A. Rizk, D. Milic, J. Zhou, B. Breton, C. Le Gouill, A. Inoue, M. Bouvier, D. Veprintsev, Michaelis-Menten quantification of ligand signalling bias applied to the promiscuous Vasopressin V2 receptor, Mol. Pharmacol. (2022) MOLPHARM-AR-2022-000497. https://doi.org/10.1124/molpharm.122.000497.

[61] W. Tan, D. Martin, J.S. Gutkind, The Gα13-Rho Signaling Axis Is Required for SDF-1-induced Migration through CXCR4, J. Biol. Chem. 281 (2006) 39542–39549. https://doi.org/10.1074/jbc.M609062200.

[62] I. Szokodi, P. Tavi, G. Földes, S. Voutilainen-Myllylä, M. Ilves, H. Tokola, S. Pikkarainen, J. Piuhola, J. Rysä, M. Tóth, H. Ruskoaho, Apelin, the Novel Endogenous Ligand of the Orphan Receptor APJ, Regulates Cardiac Contractility, Circ. Res. 91 (2002) 434–440. https://doi.org/10.1161/01.RES.0000033522.37861.69.

[63] B. Bai, Y. Jiang, X. Cai, J. Chen, Dynamics of apelin receptor/G protein coupling in living cells, Exp. Cell Res. 328 (2014) 401–409. https://doi.org/10.1016/j.yexcr.2014.08.035.

[64] K. Tran, X. Sainsily, J. Côté, D. Coquerel, P. Couvineau, S. Saibi, L. Haroune, É. Besserer-Offroy, J. Flynn-Robitaille, M. Resua Rojas, A. Murza, J.-M. Longpré, M. Auger-Messier,

O. Lesur, M. Bouvier, É. Marsault, P.-L. Boudreault, P. Sarret, Size-Reduced Macrocyclic Analogues of [Pyr $^{1}$ ]-apelin-13 Showing Negative Gα $_{12}$ Bias Still Produce Prolonged Cardiac Effects, J. Med. Chem. 65 (2022) 531–551. https://doi.org/10.1021/acs.jmedchem.1c01708.

[65] A. Novikoff, S.L. O'Brien, M. Bernecker, G. Grandl, M. Kleinert, P.J. Knerr, K. Stemmer, M. Klingenspor, A. Zeigerer, R. DiMarchi, M.H. Tschöp, B. Finan, D. Calebiro, T.D. Müller, Spatiotemporal GLP-1 and GIP receptor signaling and trafficking/recycling dynamics induced by selected receptor mono- and dual-agonists, Mol. Metab. 49 (2021) 101181. https://doi.org/10.1016/j.molmet.2021.101181.

[66] X. Zhan, L.E. Gimenez, V.V. Gurevich, B.W. Spiller, Crystal structure of arrestin-3 reveals the basis of the difference in receptor binding between two non-visual subtypes, J. Mol. Biol. 406 (2011) 467–478. https://doi.org/10.1016/j.jmb.2010.12.034.

[67] X.E. Zhou, Y. He, P.W. de Waal, X. Gao, Y. Kang, N. Van Eps, Y. Yin, K. Pal, D. Goswami, T.A. White, A. Barty, N.R. Latorraca, H.N. Chapman, W.L. Hubbell, R.O. Dror, R.C. Stevens, V. Cherezov, V.V. Gurevich, P.R. Griffin, O.P. Ernst, K. Melcher, H.E. Xu, Identification of Phosphorylation Codes for Arrestin Recruitment by G Protein-Coupled Receptors, Cell 170 (2017) 457-469.e13. https://doi.org/10.1016/j.cell.2017.07.002.

[68] E. Miess, A.B. Gondin, A. Yousuf, R. Steinborn, N. Mösslein, Y. Yang, M. Göldner, J.G. Ruland, M. Bünemann, C. Krasel, M.J. Christie, M.L. Halls, S. Schulz, M. Canals, Multisite phosphorylation is required for sustained interaction with GRKs and arrestins during rapid μ-opioid receptor desensitization, Sci. Signal. 11 (2018) eaas9609. https://doi.org/10.1126/scisignal.aas9609.

[69] S.-R. Jung, C. Kushmerick, J.B. Seo, D.-S. Koh, B. Hille, Muscarinic receptor regulates extracellular signal regulated kinase by two modes of arrestin binding, Proc. Natl. Acad. Sci. U. S. A. 114 (2017) E5579–E5588. https://doi.org/10.1073/pnas.1700331114.

[70] M.D. Richardson, A.M. Balius, K. Yamaguchi, E.R. Freilich, L.S. Barak, M.M. Kwatra, Human substance P receptor lacking the C-terminal domain remains competent to desensitize and internalize, J. Neurochem. 84 (2003) 854–863. https://doi.org/10.1046/j.1471-4159.2003.01577.x.

[71] V.R. Jala, W.-H. Shao, B. Haribabu, Phosphorylation-independent beta-arrestin translocation and internalization of leukotriene B4 receptors, J. Biol. Chem. 280 (2005) 4880–4887. https://doi.org/10.1074/jbc.M409821200.

# FIGURES

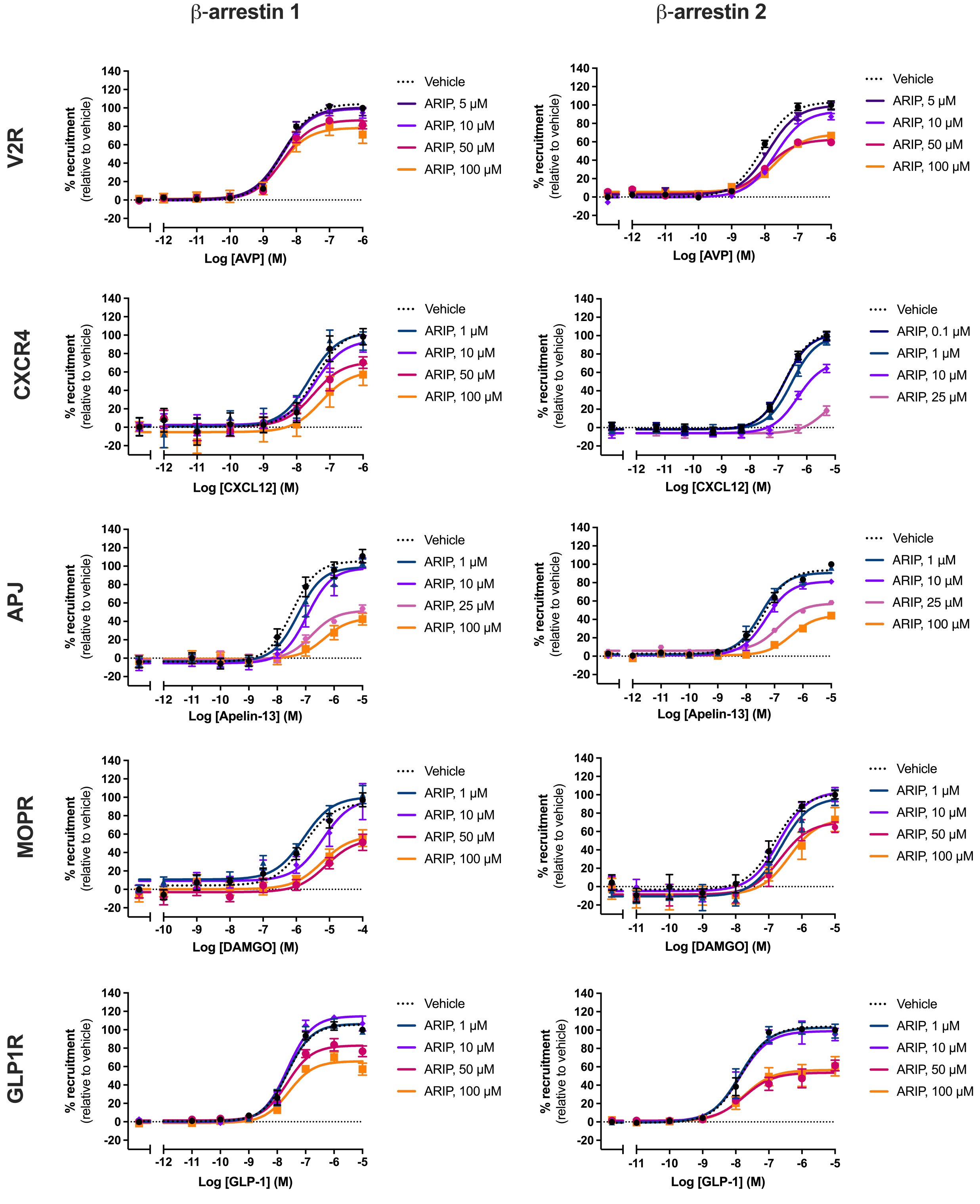

**Figure 1. ARIP inhibits agonist-induced β-arrestin 1 and 2 recruitment at select 7TMRs.** β-arrestin 1 and 2 recruitment to 7TM receptors V2R, CXCR4, APJ, MOPR and GLP1R in transiently transfected HEK293 cells was monitored via $BRET^2$ assays. Cells were treated with fixed concentrations of ARIP (1-100 μM) for 20 min prior to stimulation with increasing agonist concentrations (arginine-vasopressin (AVP), CXCL12, apelin-13, DAMGO and GLP-1, respectively). Plates were read 20 min after agonist stimulation. Two-way ANOVA analyses revealed that ARIP concentration significantly affected β-arrestin recruitment at all receptors tested ($p<0.01$). Data represent the mean ± SEM of 3 independent experiments, each performed in duplicate.

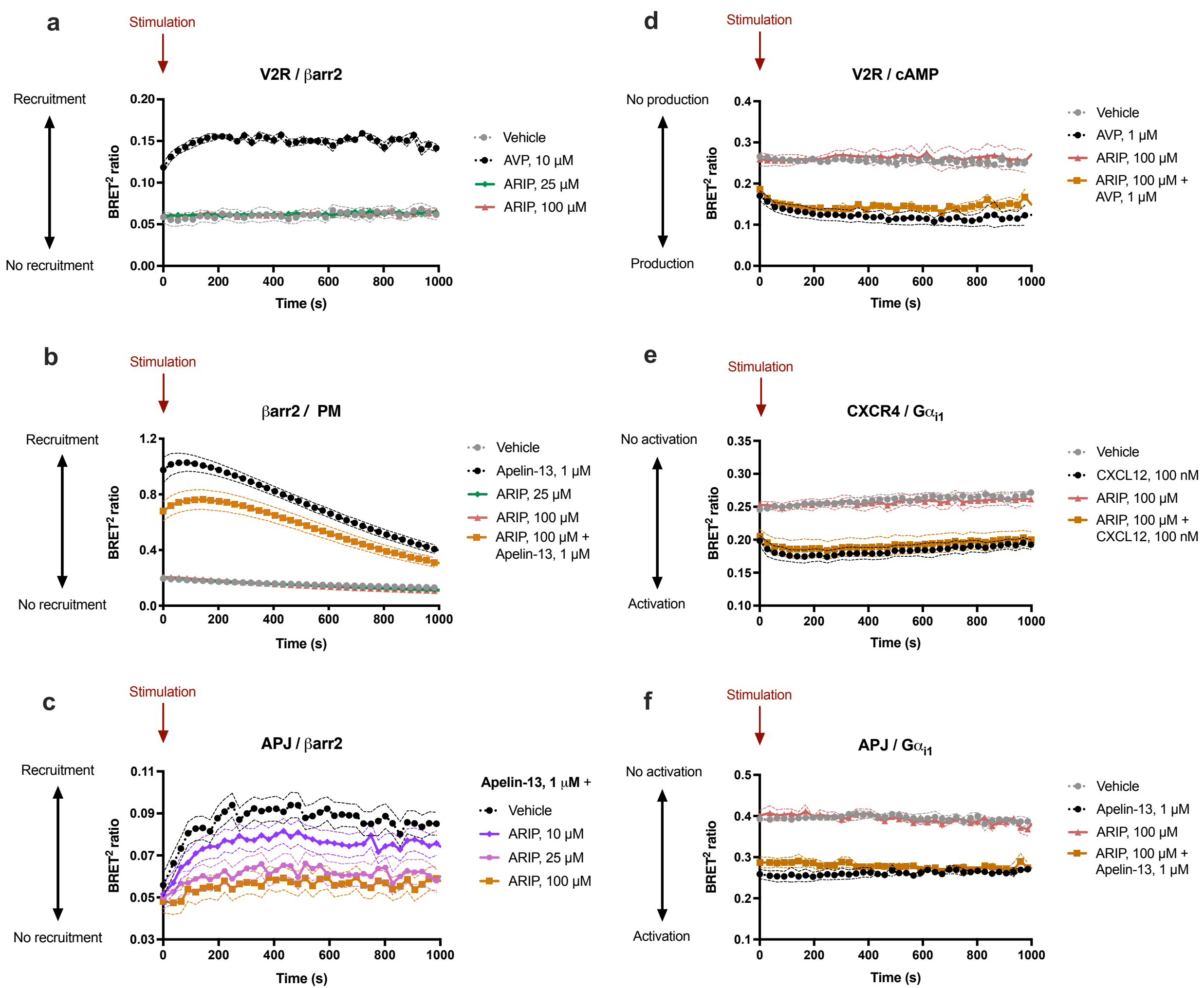

**Figure 2. ARIP does not promote β-arrestin recruitment without receptor activation, nor does it affect ligand-induced G protein activation at select 7TMRs.** BRET² time-course experiments in transiently transfected HEK293 cells monitoring: (**a**) β-arrestin 2 recruitment at V2R (V2R-RlucII, β-arrestin 2-GFP10), treated with AVP (10 µM) or ARIP (25, 100 µM) alone. (**b**) β-arrestin 2 recruitment to the plasma membrane (APJ, β-arrestin 2-RlucII, rGFP-CAAX), treated with Apelin-13 (1 µM) and/or ARIP (25, 100 µM). (**c**) apelin-13-induced β-arrestin 2 recruitment at APJ, and pre-treatment with ARIP (0, 10, 25, 100 µM) decreases β-arrestin 2 recruitment to APJ in a concentration-dependent manner as analyzed by two-way ANOVA ($p<0.01$). (**d-f**) Agonist-induced cAMP production and $G\alpha_{i1}$ activation at V2R, APJ, and CXCR4, with or without pre-treatment with ARIP (100 µM), and ARIP (100 µM) tested alone. Data

represent the mean ± SEM of 3 independent experiments, each performed in triplicate. Readings were taken at 30 s intervals.

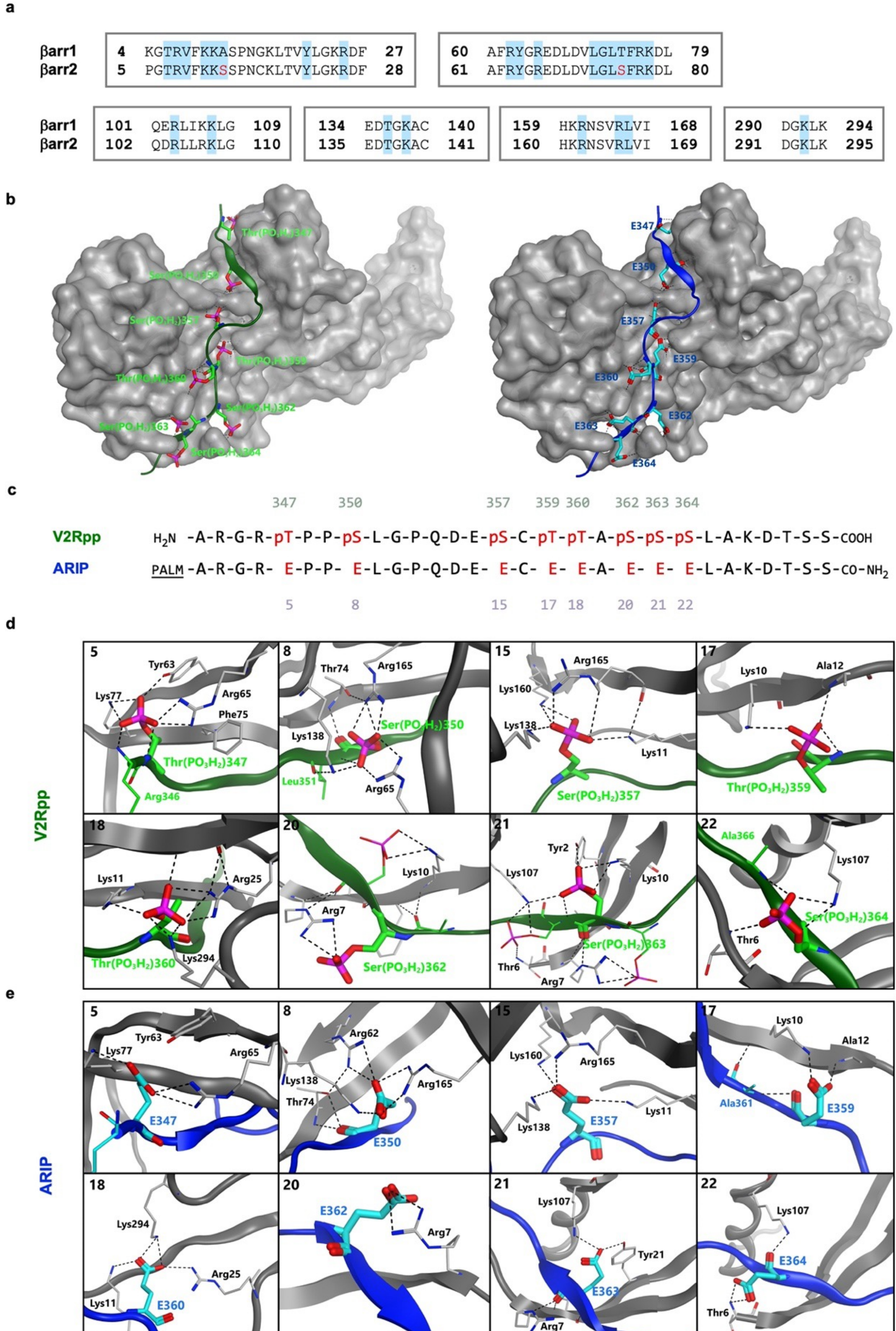

**Figure 3. Molecular modelling suggests that ARIP binds β-arrestin at the same site as V2Rpp, but that key interactions do not occur due to missing phosphoryl groups.** (**a**) Key sections of a βarr1/βarr2 alignment, performed using Clustal Omega. Sequences were defined according to Uniprot database (βarr1: P49407, βarr2: P32121). Residues highlighted in blue were identified as V2Rpp/βarr binding sites, according to the crystal structure (PDB: 4JQI). βarr2 residues that differ from βarr1 at binding sites are shown in red. V2Rpp-binding residues are highly conserved between βarr1 and βarr2. (**b**) 3D contact maps of V2Rpp (green) or ARIP (blue) on the surface of βarr1. (**c**) Amino acid sequences of the V2Rpp peptide and ARIP. Phosphosites are numbered according to their position within the V2R receptor (above, grey-green) and their position within the peptidic portion of ARIP (below, grey-purple). (**d**) Interactions between V2Rpp (green) or (**e**) ARIP (blue) and βarr1 at each of the 8 phosphosites (positions 5, 8, 15, 17, 18, 20, 21, and 22).

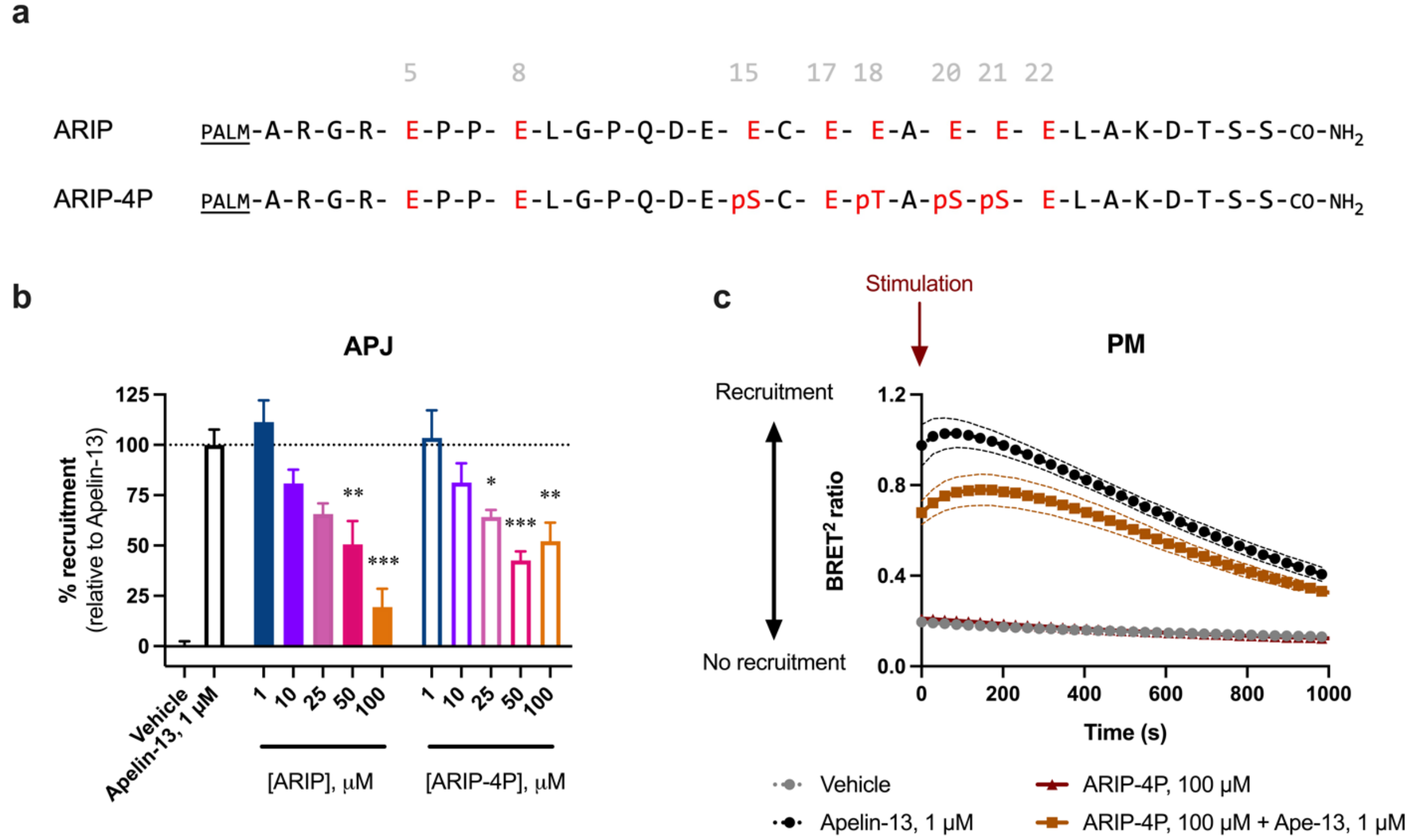

**Figure 4. ARIP-4P inhibits agonist-induced β-arrestin recruitment similarly to ARIP.** (**a**) Peptide sequences of the lipopeptide β-arrestin inhibitors, ARIP and ARIP-4P. For ARIP-4P, phosphorylated serines and threonines were inserted at sites **15** (S357), **18** (T360), **20** (S362), and **21** (S363). (**b, c**) β-arrestin 2 recruitment to APJ (**b**) and to the plasma membrane (**c**) in transiently transfected HEK293 cells, monitored by $BRET^2$. Cells were treated with fixed concentrations (0, 1, 10, 25, 50, 100 μM (**b**) or 0, 100 μM (**c**) of ARIP or ARIP-4P 20 min prior to stimulation with 1 μM of Apelin-13 agonist. Plates were read 10 min after agonist stimulation (**b**). Note that the β-arrestin 2-RlucII and rGFP-CAAX biosensors were co-transfected with APJ as a control (**c**). A one-way ANOVA with Dunnett's correction for multiple comparisons was performed. Asterisks (*) denote a statistically significant difference from the agonist condition (Apelin-13, 1 μM). One symbol, $p$=0.05; two, $p$=0.01; three, $p$=0.001. Data represent mean ± SEM of 3 independent experiments, each performed in duplicate.

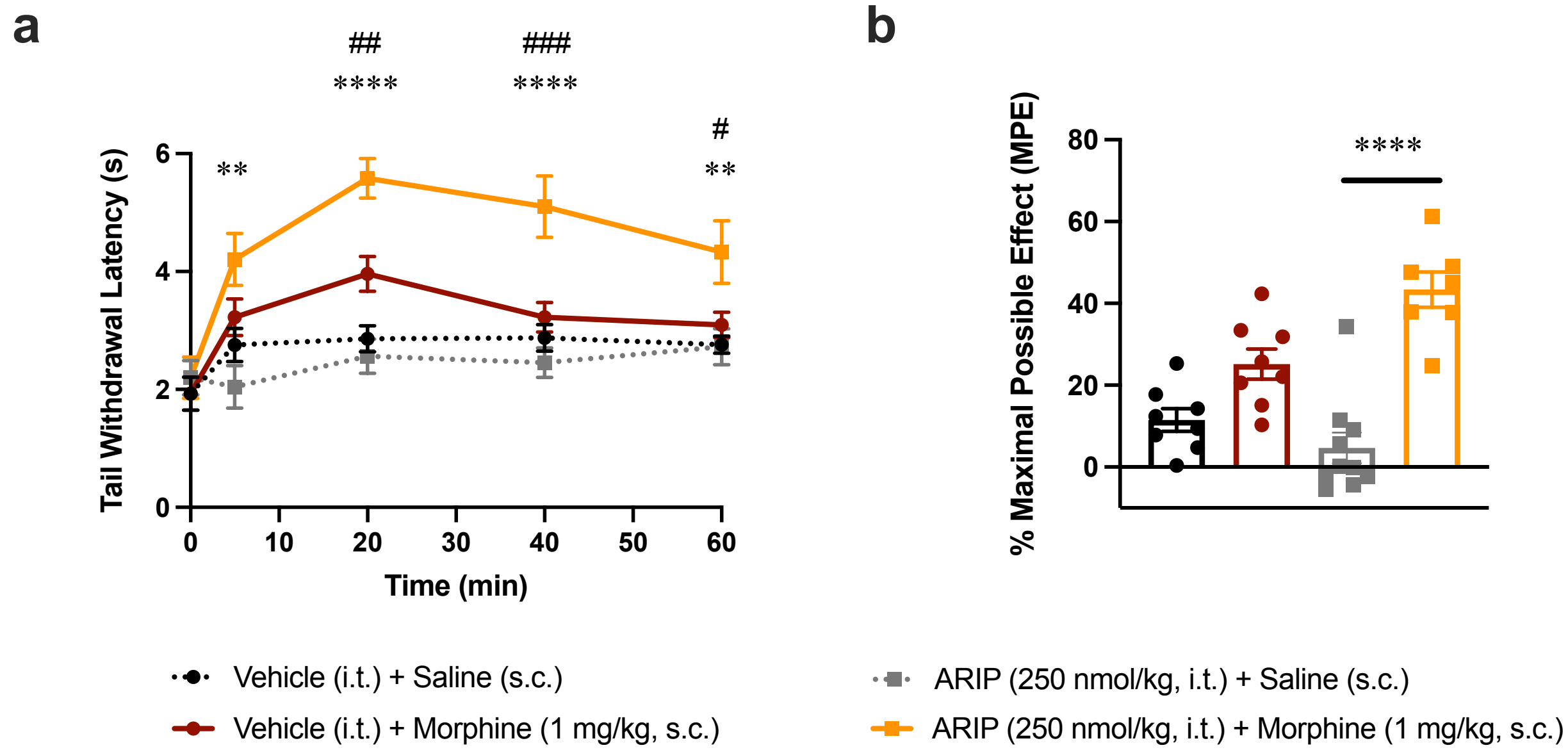

**Figure 5. Co-administration of ARIP enhances morphine-induced antinociception in a rat model of acute thermal pain.** (**a**) Tail withdrawal latencies of male Sprague-Dawley rats co-injected with vehicle (saline containing 50% DMSO and 20% PEG; i.t.) or ARIP (250 nmol/kg; i.t.), and saline (s.c.) or morphine (1 mg/kg; s.c.). (**b**) Maximal Possible Effect (%) 20 minutes after morphine injection, calculated considering a cut-off of 10 s. A two-way ANOVA with Tukey's correction for multiple comparisons was performed in (a), and a Kruskal-Wallis with Dunn's correction for multiple comparisons was performed in (b). Asterisks (*) denote statistical differences between the "ARIP + Morphine" and "ARIP + Saline" groups. Sharps (#) indicate statistical differences between the "ARIP + Morphine" and "Vehicle + Morphine" groups. One symbol, $p$=0.05; two, $p$=0.01; three, $p$=0.001; four, $p$=0.0001. Data represent mean ± SEM, $n$=7 rats.

## TABLES

**Table 1. Peptide sequence of Arrestin Recruitment Inhibitory Peptide (ARIP).** ARIP is an N-terminally palmitoylated analogue of V2Rpp, a phosphorylated peptide derived from the C-terminal A343 to S371 amino acids of V2R. For ARIP, the 8 phosphorylated residues at positions **5** (T347), **8** (S350), **15** (S357), **17** (T359), **18** (T360), **20** (S362), **21** (S363), and **22** (S364) have been replaced by phosphomimetic Glu (E) residues. An N-terminal palmitate residue was added to enable ARIP to be anchored to the membrane and penetrate into the cell.

| **Compound** | **Peptide sequence** |
|---|---|
| V2Rpp | $H_2N$ -A-R-G-R-pT-P-P-pS-L-G-P-Q-D-E-pS-C-pT-pT-A-pS-pS-pS-L-A-K-D-T-S-S- COOH |
| ARIP | PALM -A-R-G-R- E-P-P- E-L-G-P-Q-D-E- E-C- E- E-A- E- E- E-L-A-K-D-T-S-S- $CONH_2$ |

**Table 2. Effect of ARIP on agonists' potency to recruit β-arrestins 1 and 2.**

| | $EC_{50}$ ± SEM, (nM) | | | | | | | | | |
|---|---|---|---|---|---|---|---|---|---|---|
| | **V2R** | | **CXCR4** | | **APJ** | | **MOR** | | **GLP1R** | |
| **[ARIP], (μM)** | βarr1 | βarr2 | βarr1 | βarr2 | βarr1 | βarr2 | βarr1 | βarr2 | βarr1 | βarr2 |
| **0** | 4.1 ± 0.27 | 8.3 ± 0.47 | 40 ± 11 | 170 ± 20 | 34 ± 6.2 | 46 ± 3.6 | 1700 ± 300 | 150 ± 26 | 23 ± 2.2 | 15 ± 1.2 |
| **0.1** | — | — | — | 170 ± 20 | — | — | — | — | — | — |
| **1** | — | — | 24 ± 6.8 | 300 ± 35 | 50 ± 11 | 34 ± 3.5 | 1700 ± 320 | 230 ± 58 | 23 ± 2.6 | 13 ± 2.2 |
| **5** | 3.7 ± 0.33 | **13.3 ± 0.76** | — | — | — | — | — | — | — | — |
| **10** | 3.8 ± 0.39 | **21 ± 1.4** | 36 ± 9.6 | 480 ± 76 | **110 ± 20** | 51 ± 4.9 | **6000 ± 2600** | 210 ± 28 | 23 ± 2.9 | 14 ± 2.5 |
| **25** | — | — | 50 ± 14 | N/C | **160 ± 43** | **140 ± 19** | — | — | — | — |
| **50** | 4.0 ± 0.43 | 12 ± 1.1 | 32 ± 9.5 | — | — | — | **8000 ± 2600** | 230 ± 71 | 22 ± 3.4 | 18 ± 4.6 |
| **100** | 3.4 ± 0.82 | **21 ± 1.9** | 60 ± 28 | — | **500 ± 210** | **400 ± 62** | **6000 ± 2000** | **400 ± 180** | 25 ± 4.0 | 18 ± 4.6 |

— : not tested. N/C : not converged – unable to generate a reliable $EC_{50}$. Values in bold are statistically different from agonist condition.

**Table 3. Effect of ARIP on agonists' efficacy to recruit β-arrestins 1 and 2.**

| | $E_{max}$ ± SEM, % | | | | | | | | | |
|---|---|---|---|---|---|---|---|---|---|---|
| | **V2R** | | **CXCR4** | | **APJ** | | **MOPR** | | **GLP1R** | |
| **[ARIP], (μM)** | βarr1 | βarr2 | βarr1 | βarr2 | βarr1 | βarr2 | βarr1 | βarr2 | βarr1 | βarr2 |
| **0** | 100 ± 3.2 | 100 ± 4.4 | 100 ± 8.8 | 100 ± 7.6 | 100 ± 5.8 | 100 ± 2.5 | 100 ± 7.6 | 100 ± 4.4 | 100 ± 2.8 | 100 ± 3.2 |
| **0.1** | — | — | — | 99 ± 4.8 | — | — | — | — | — | — |
| **1** | — | — | 90 ± 10 | 94 ± 4.6 | 103 ± 4.3 | 96 ± 0.96 | 103 ± 9.8 | 97 ± 8.7 | 107 ± 5.0 | 102 ± 6.4 |
| **5** | 99 ± 4.5 | 99 ± 4.1 | — | — | — | — | — | — | — | — |
| **10** | 97 ± 5.6 | 87 ± 3.4 | 90 ± 8.9 | **60 ± 20** | 103 ± 5.8 | **81 ± 1.4** | 100 ± 20 | 100 ± 7.7 | 113 ± 3.2 | 100 ± 12 |
| **25** | — | — | 70 ± 14 | **18 ± 5.0** | **54 ± 4.0** | **58 ± 0.66** | — | — | — | — |
| **50** | 86 ± 3.6 | **59 ± 3.0** | 70 ± 6.3 | — | — | — | **51 ± 8.9** | **65 ± 5.7** | 84 ± 6.6 | **61 ± 5.8** |
| **100** | 80 ± 11 | **67 ± 3.0** | **60 ± 12** | — | **42 ± 6.5** | **44 ± 2.1** | 56 ± 9.1 | 70 ± 13 | **57 ± 6.6** | **60 ± 10** |

— : not tested. Values in bold are statistically different from agonist condition.

**Table 4. Effect of ARIP and ARIP-4P on agonists' efficacy to recruit β-arrestin 2**

| | Mean ± SEM, % | | | | | |
|---|---|---|---|---|---|---|
| | APJ | | | | | |
| **[Inhibitor], (μM)** | **0** | **1** | **10** | **25** | **50** | **100** |
| **ARIP** | 100 ± 7.3 | 110 ± 11 | 81 ± 6.8 | 66 ± 5.3 | **50 ± 12** | **19 ± 9.1** |
| **ARIP-4P** | 100 ± 7.3 | 100 ± 14 | 81 ± 9.5 | **64 ± 3.5** | **43 ± 4.5** | **52 ± 9.3** |

Values in bold are statistically different from agonist condition.

## SUPPLEMENTARY INFORMATION

# A lipidated peptide derived from the C-terminal tail of the vasopressin 2 receptor shows promise as a new β-arrestin inhibitor

Rebecca L. Brouillette[a,b], Christine E. Mona[c,d], Michael Desgagné[a,b], Malihe Hassanzedeh[a,b], Émile Breault[a,b], Frédérique Lussier[a,b], Karine Belleville[a,b], Jean-Michel Longpré[a,b], Michel Grandbois[a,b], Pierre-Luc Boudreault[a,b,e], Élie Besserer-Offroy[a,e,f,g,#,*], Philippe Sarret[a,b,e,#*]

[a]Department of Pharmacology-Physiology, Faculty of Medicine and Health Sciences, Université de Sherbrooke, Sherbrooke, QC, Canada.

[b]Institut de Pharmacologie de Sherbrooke, Université de Sherbrooke, Sherbrooke, QC, Canada.

[c]Ahmanson Translational Theranostics Division, Department of Molecular and Medical Pharmacology, David Geffen School of Medicine, University of California-Los Angeles, Los Angeles, CA, USA.

[d]Jonsson Comprehensive Cancer Center, UCLA Health, Los Angeles, CA, USA.

[e]RECITAL International Partnership Lab, Université de Caen-Normandie, Caen, France & Université de Sherbrooke, Sherbrooke, QC, Canada.

[f]Université de Caen Normandie, INSERM U1086 – Anticipe, Normandie Université, Caen, France.

[g]Baclesse Comprehensive Cancer Center, UNICANCER, Caen, France.

[#]Lead Authors

***Co-Corresponding Authors**

**Philippe Sarret, Ph.D.**
Department of Pharmacology-Physiology
Faculty of Medicine and Health Sciences
Université de Sherbrooke
3001, 12th Avenue North
Sherbrooke, QC, Canada, J1H 5N4
Ph.: +1 (819) 821-8000, Ext. 72554
Philippe.Sarret@USherbrooke.ca

**Elie Besserer-Offroy, Ph.D.**
Inserm U1086 – Anticipe
Baclesse Comprehensive Cancer Center
Université de Caen-Normandie
3 avenue du Général Harris
14076 Caen Cedex 5, France
Ph.: +33 2 31 45 50 70
Elie.Besserer-Offroy@inserm.fr

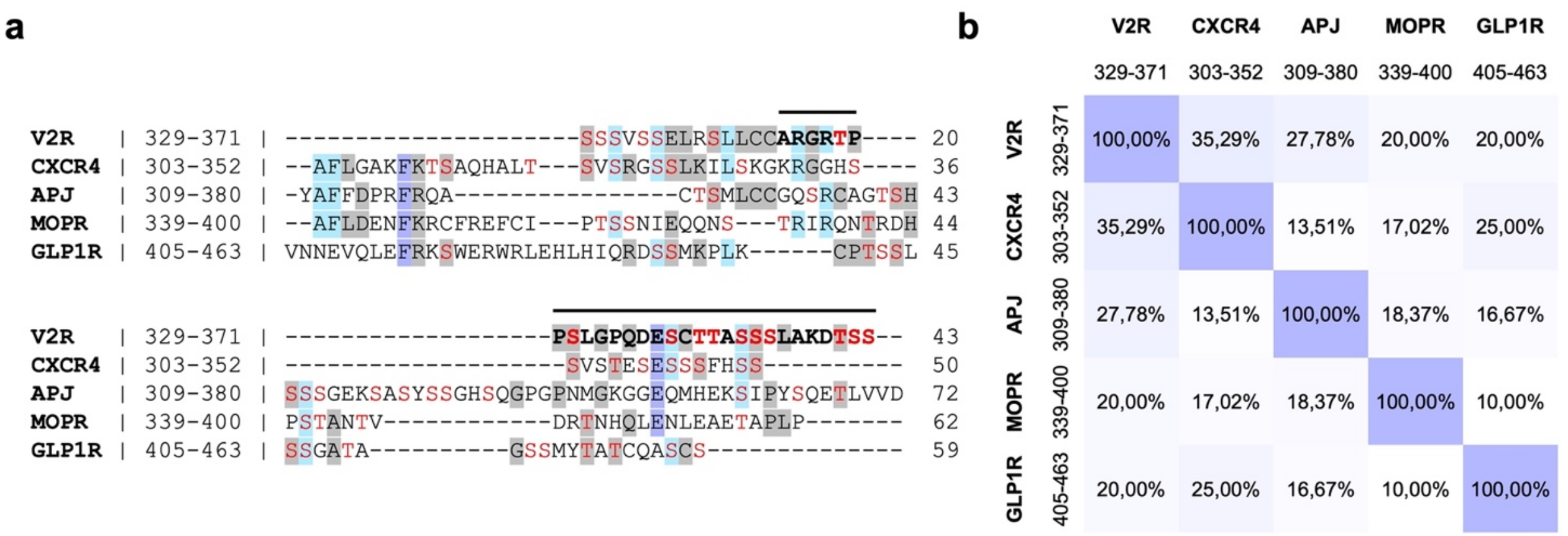

**Supplementary Figure S1. Alignment of 7TMR C-terminal domains**. (a) The C-terminal domains of tested 7TMRs were aligned using the Clustal Omega Program at Uniprot.org. Sequences were defined according to their Uniprot IDs: P30518 (V2R), P61073 (CXCR4), P35414 (APJ), P35372 (MOPR), and P43220 (GLP1R). Residues A343 to S371 of V2R are shown in bold. Phosphorylatable Ser and Thr residues are shown in red. At each position, identical residues between two receptor sequences are highlighted in grey; three, in light blue; and four, in light purple. (b) Identity matrix showing % identity between the C-terminal sequences of 7TMRs.

**Supplementary Table S1. Chemical characterisation of ARIP and ARIP-4P**

| Compound | Peptide sequence | Molecular formula | % Purity (290 nm) | Observed ion | Calculated m/z | HRMS |
|---|---|---|---|---|---|---|
| ARIP | Palm-A-R-G-R-E-P-P-E-L-G-P-Q-D-E-E-C-E-E-A-E-E-E-L-A-K-D-T-S-S-$NH_2$ | $C_{145}H_{236}N_{38}O_{56}S$ | 97 | $[M+3H]^{3+}$ | 1146.8909 | 1146.8902 |
| ARIP-4P | Palm-A-R-G-R-E-P-P-E-L-G-P-Q-D-E-pS-C-E-pT-A-pS-pS-E-L-A-K-D-T-S-S-$NH_2$ | $C_{138}H_{234}N_{38}O_{64}P_4S$ | 97 | $[M+3H]^{3+}$ | 1202.1705 | 1202.1709 |

**Supplementary Table S2: Effect of ARIP on agonists' potency to recruit β-arrestins 1 and 2 – Statistical differences.**

| | ***P* values ($pEC_{50}$)** | | | | | | | | | |
|---|---|---|---|---|---|---|---|---|---|---|
| | **V2R** | | **CXCR4** | | **APJ** | | **MOR** | | **GLP1R** | |
| **[ARIP], (μM)** | βarr1 | βarr2 | βarr1 | βarr2 | βarr1 | βarr2 | βarr1 | βarr2 | βarr1 | βarr2 |
| **0** | N/A | N/A | N/A | N/A | N/A | N/A | N/A | N/A | N/A | N/A |
| **0.1** | — | — | — | 0.9999 | — | — | — | — | — | — |
| **1** | — | — | 0.8483 | 0.8905 | 0.5510 | 0.1517 | 0.9999 | 0.5658 | 0.9999 | 0.9871 |
| **5** | 0.9397 | **0.0162** | — | — | — | — | — | — | — | — |
| **10** | 0.9756 | **<0.0001** | 0.9999 | 0.5218 | **0.0272** | 0.9071 | **0.0123** | 0.7869 | 0.9999 | 0.9934 |
| **25** | — | — | 0.9279 | N/C | **0.0058** | **<0.0001** | — | — | — | — |
| **50** | 0.9991 | 0.0937 | 0.9997 | — | — | — | **<0.0001** | 0.5910 | 0.9991 | 0.9099 |
| **100** | 0.7595 | **<0.0001** | 0.7907 | — | **0.0001** | **<0.0001** | **0.0154** | **0.0324** | 0.9645 | 0.9099 |

— : not tested. N/A: not applicable (reference condition). N/C: not converged. Values in bold represent statistically significant *p* values.

**Supplementary Table S3: Effect of ARIP on agonists' potency to recruit β-arrestins 1 and 2 – Fold increase.**

| | A' / A; ($EC_{50}$ 'agonist + ARIP')/ ($EC_{50}$ 'agonist') | | | | | | | | | |
|---|---|---|---|---|---|---|---|---|---|---|
| | V2R | | CXCR4 | | APJ | | MOR | | GLP1R | |
| **[ARIP], (μM)** | βarr1 | βarr2 | βarr1 | βarr2 | βarr1 | βarr2 | βarr1 | βarr2 | βarr1 | βarr2 |
| **0** | 1.0 | 1.0 | 1.0 | 1.0 | 1.0 | 1.0 | 1.0 | 1.0 | 1.0 | 1.0 |
| **0.1** | — | — | — | 1.0 ± 0.12 | — | — | — | — | — | — |
| **1** | — | — | 0.7 ± 0.19 | 1.7 ± 0.21 | 1.6 ± 0.32 | 0.72 ± 0.08 | 1.0 ±0.19 | 1.5 ± 0.38 | 1.0 ± 0.11 | 0.9 ± 0.15 |
| **5** | 0.90 ± 0.08 | **1.61 ± 0.1** | — | — | — | — | — | — | — | — |
| **10** | 0.92 ± 0.1 | **2.5 ± 0.17** | 1.0 ± 0.26 | 2.8 ± 0.44 | **3.3 ± 0.60** | **1.1 ± 0.11** | **4 ± 1.2** | 1.3 ± 0.18 | 1.0 ± 0.13 | 0.9 ± 0.17 |
| **25** | — | — | 1.4 ± 0.40 | N/D | **5 ± 1.3** | **3.0 ± 0.41** | — | — | — | — |
| **50** | 1.0 ± 0.10 | 1.4 ± 0.13 | 0.9 ± 0.26 | — | — | — | **5 ± 1.6** | 1.5 ± 0.46 | 1.0 ± 0.14 | 1.2 ± 0.31 |
| **100** | 0.8 ± 0.20 | **2.6 ± 0.23** | 1.3 ± 0.80 | — | **13 ± 6.1** | **9 ± 1.3** | **4 ± 1.2** | **3 ± 1.2** | 1.1 ± 0.17 | 1.2 ± 0.31 |

— : not tested. N/D: not determined. Values in bold are statistically different from agonist condition.

**Supplementary Table S4. Effect of ARIP on agonists' efficacy to recruit β-arrestins 1 and 2 – Statistical differences.**

| | ***P* values ($E_{max}$)** | | | | | | | | | |
|---|---|---|---|---|---|---|---|---|---|---|
| | **V2R** | | **CXCR4** | | **APJ** | | **MOR** | | **GLP1R** | |
| **[ARIP], (μM)** | βarr1 | βarr2 | βarr1 | βarr2 | βarr1 | βarr2 | βarr1 | βarr2 | βarr1 | βarr2 |
| **0** | N/A | N/A | N/A | N/A | N/A | N/A | N/A | N/A | N/A | N/A |
| **0.1** | — | — | — | 0.9997 | — | — | — | — | — | — |
| **1** | — | — | 0.9988 | 0.8272 | 0.7580 | 0.4331 | 0.9908 | 0.9965 | 0.7050 | 0.9997 |
| **5** | 0.9880 | 0.9959 | — | — | — | — | — | — | — | — |
| **10** | 0.9481 | 0.0821 | 0.9753 | **0.0011** | 0.7563 | **<0.0001** | 0.9998 | 0.9999 | 0.2347 | 0.9982 |
| **25** | — | — | 0.3616 | **<0.0001** | **0.0001** | **<0.0001** | — | — | — | — |
| **50** | 0.2472 | **<0.0001** | 0.2385 | — | — | — | **0.0476** | **0.0236** | 0.1181 | **0.0207** |
| **100** | 0.0852 | **<0.0001** | **0.0455** | — | **<0.0001** | **<0.0001** | 0.0787 | 0.0955 | **0.0019** | **0.0067** |

— : not tested. N/A: not applicable (reference condition). Values in bold represent statistically significant *p* values.

**Supplementary Table S5. Effect of ARIP and ARIP-4P on agonists' efficacy to recruit β-arrestin 2 – Statistical differences.**

| | | | | | | |
|---|---|---|---|---|---|---|
| ***P* values** | | | | | | |
| **APJ** | | | | | | |
| **[Inhibitor], (μM)** | **0** | **1** | **10** | **25** | **50** | **100** |
| **ARIP** | N/A | 0.9516 | 0.5731 | 0.0638 | **0.0036** | < **0.0001** |
| **ARIP-4P** | N/A | > 0.9999 | 0.5975 | **0.0483** | **0.0007** | **0.0049** |

N/A: not applicable (reference condition). Values in bold represent statistically significant *p* values.

**Supplementary Spectrum S1: UPLC characterisation of ARIP**

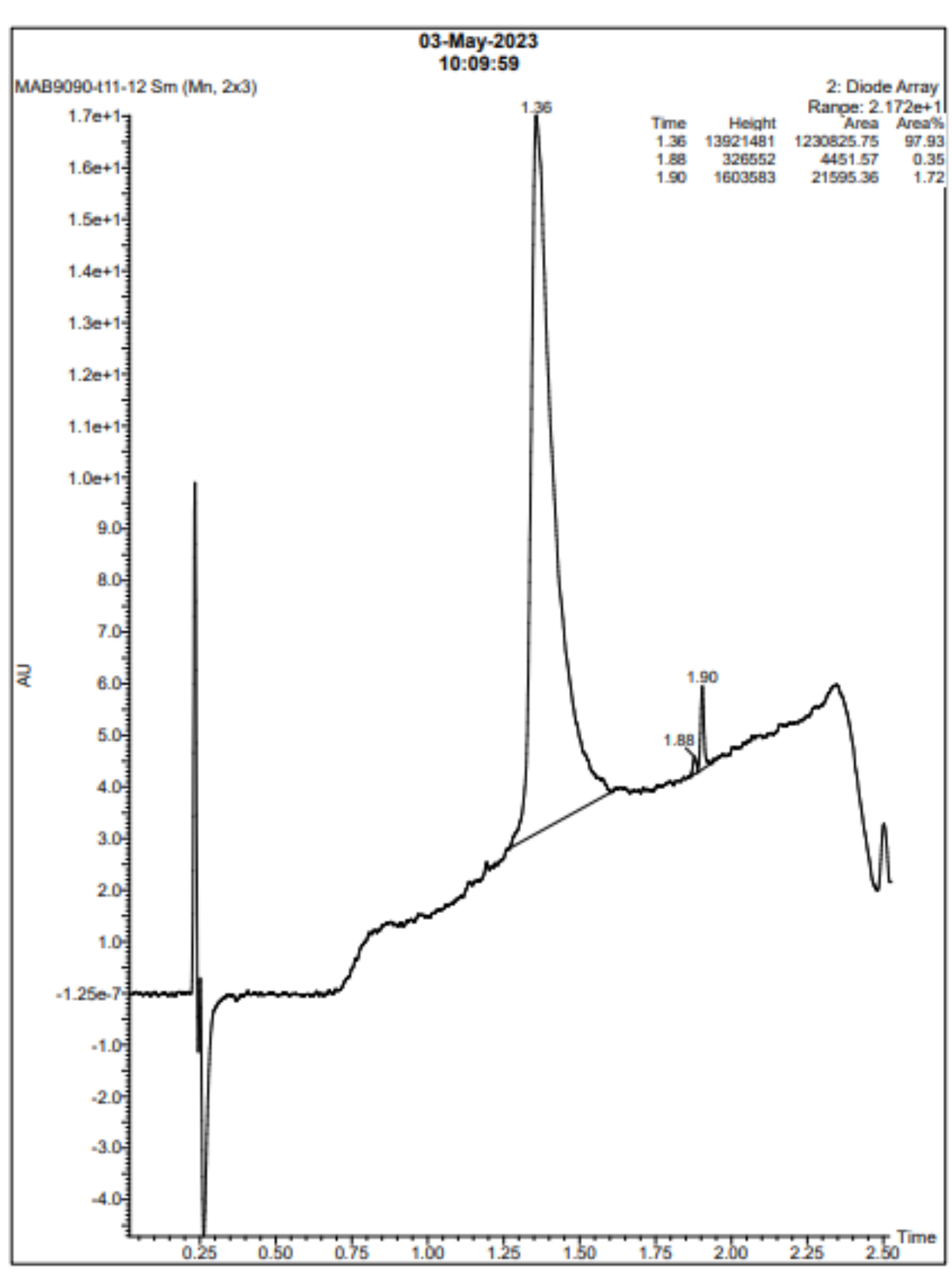

**Supplementary Spectrum S2: MS characterisation of ARIP**

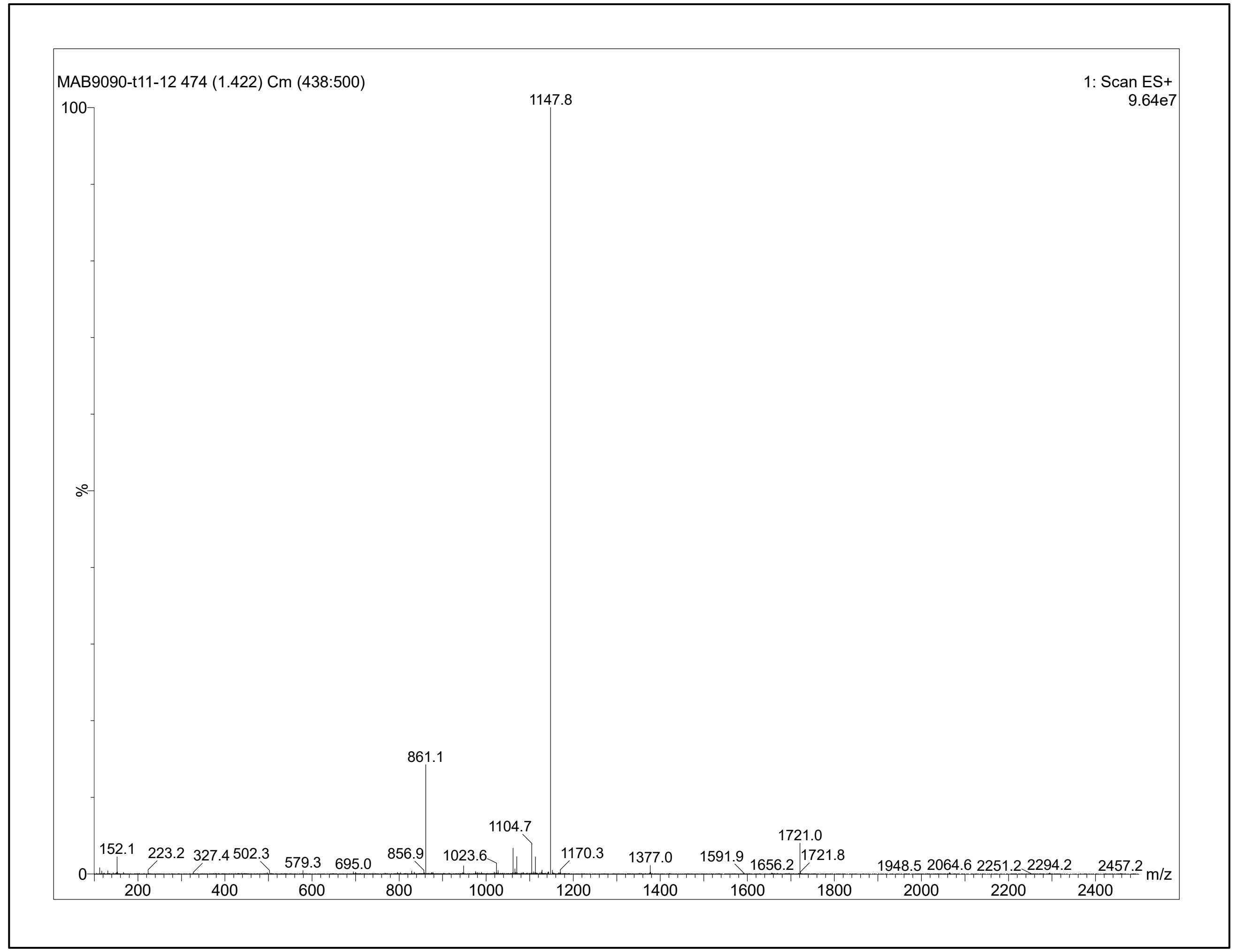

**Supplementary Spectrum S3: UPLC characterisation of ARIP-4P**

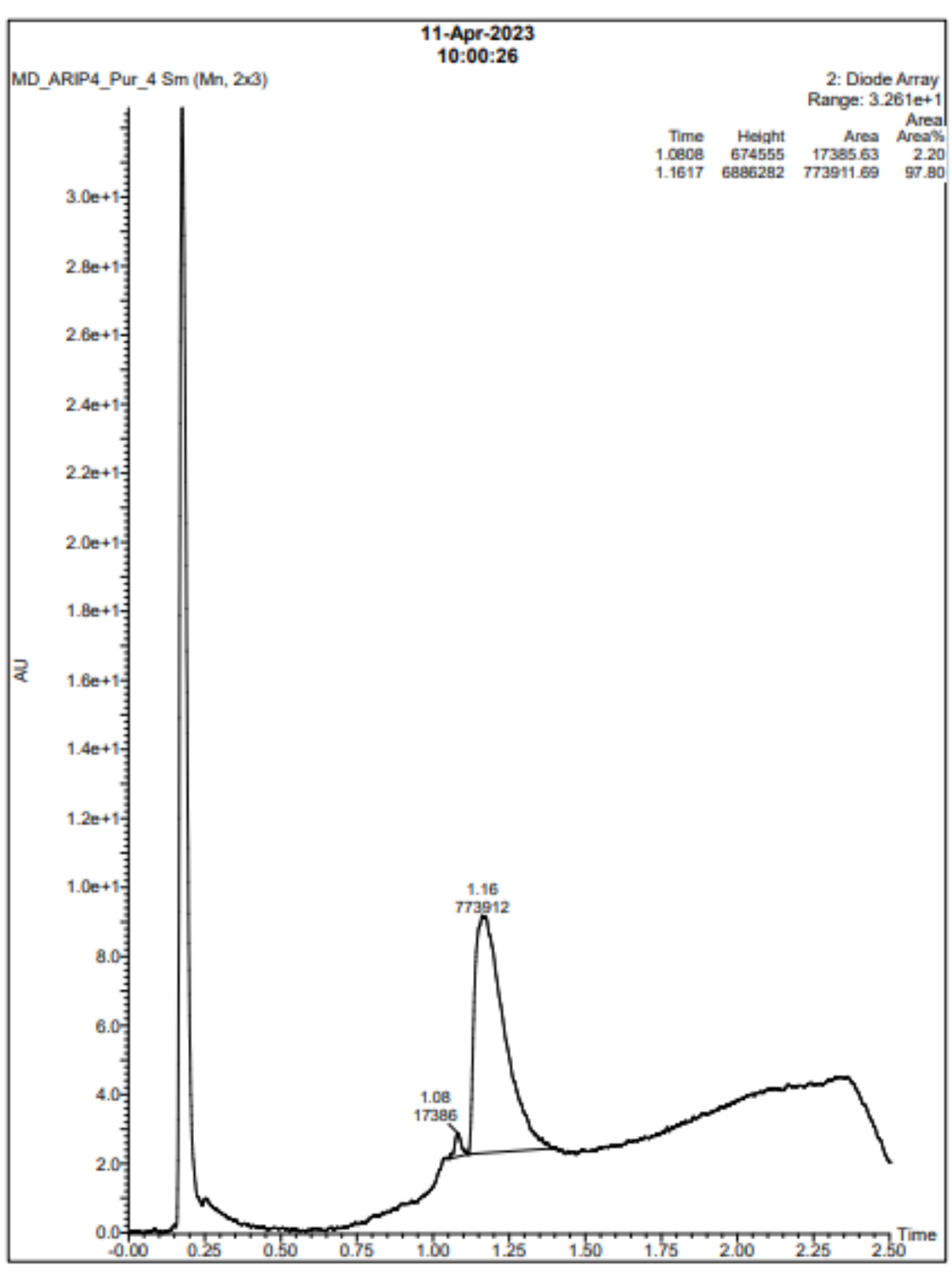

**Supplementary Spectrum S4: MS characterisation of ARIP-4P**

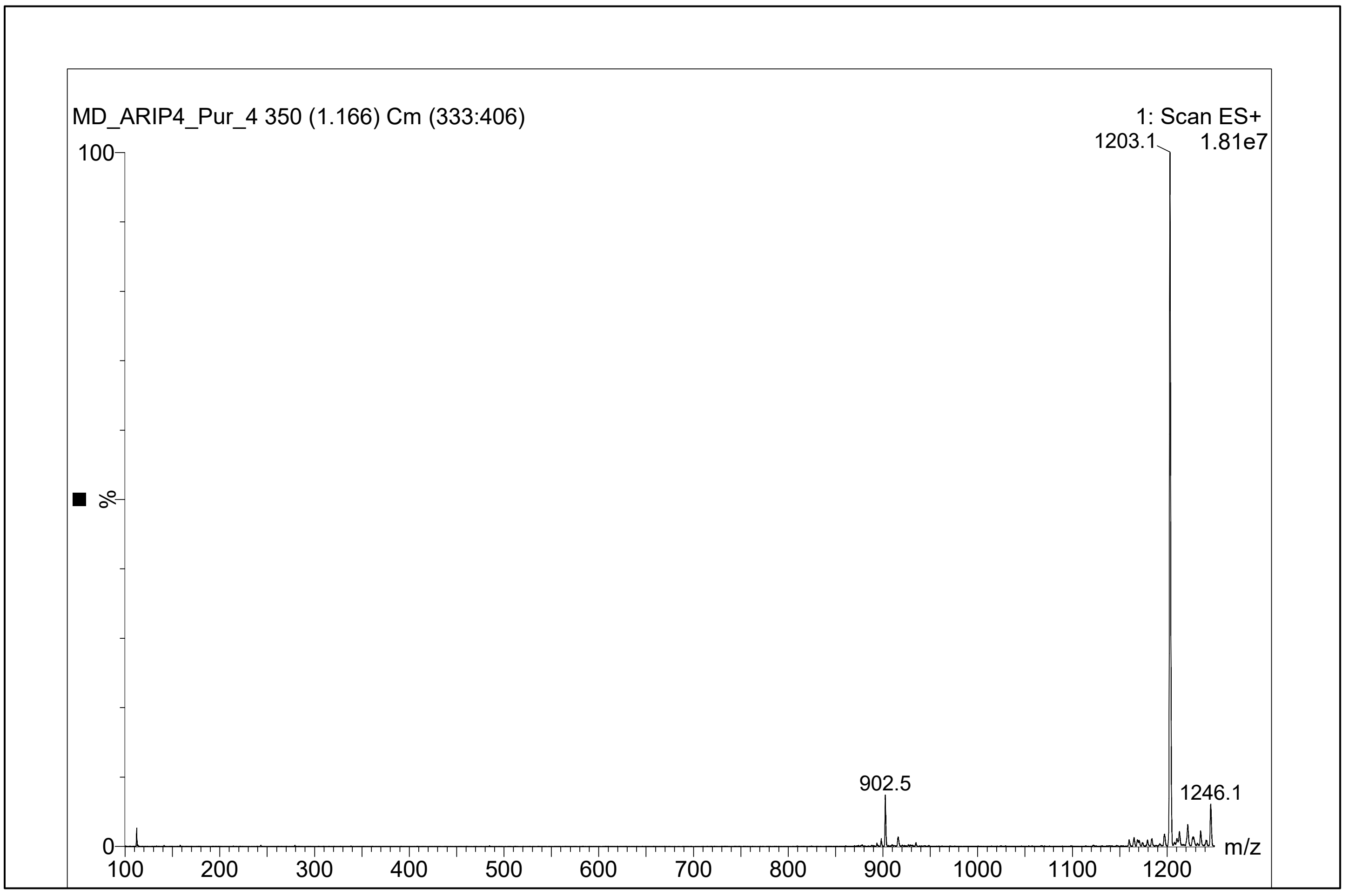